\documentclass[sigconf, anonymous=false,dvipsnames]{acmart}

\AtBeginDocument{%
  }

\usepackage[percent]{overpic}
\usepackage{subcaption} 

\copyrightyear{2022} 
\acmYear{2022} 
\setcopyright{acmlicensed}\acmConference[ICAIF '22]{3rd ACM International Conference on AI in Finance}{November 2--4, 2022}{New York, NY, USA}
\acmBooktitle{3rd ACM International Conference on AI in Finance (ICAIF '22), November 2--4, 2022, New York, NY, USA}
\acmPrice{15.00}
\acmDOI{10.1145/3533271.3561753}
\acmISBN{978-1-4503-9376-8/22/10}

\begin{document}

\title{Learning to simulate realistic limit order book markets from data as a World Agent}

\author{Andrea Coletta}
\orcid{1234-5678-9012}
\email{andrea.coletta@jpmchase.com}
\affiliation{%
  \institution{J.P. Morgan AI Research}
  \city{New York}
  \country{USA}}

\author{Aymeric Moulin}
\authornote{The research work was carried out when Aymeric Moulin was employed at J.P. Morgan AI Research.}
\email{amoulin@bamfunds.com}
\affiliation{%
  \institution{Balyasny Asset Management, L.P.}
  \city{New York}
  \country{USA}}

\author{Svitlana Vyetrenko}
\email{svitlana.s.vyetrenko@jpmchase.com}
\affiliation{%
  \institution{J.P. Morgan AI Research}
  \city{New York}
  \country{USA}}
  
\author{Tucker Balch}
\email{tucker.balch@jpmchase.com}
\affiliation{%
  \institution{J.P. Morgan AI Research}
  \city{New York}
  \country{USA}}

\renewcommand{\shortauthors}{Coletta et al.}


\begin{abstract}
Multi-agent market simulators usually require careful calibration to emulate real markets, which includes the number and the type of agents. 
%
Poorly calibrated simulators can lead to misleading conclusions, potentially causing severe loss when employed by investment banks, hedge funds, and traders to study and evaluate trading strategies.
%
In this paper, we propose a world model simulator that accurately emulates a limit order book market -- it requires no agent calibration but rather learns the simulated market behavior directly from historical data. 
%
Traditional approaches fail short to learn and calibrate trader population, as historical labeled data with details on each individual trader strategy is not publicly available.
%
Our approach proposes to learn a unique "world" agent from historical data. It is intended to emulate the overall trader population, without the need of making assumptions about individual market agent strategies.  
%
We implement our world agent simulator models as a Conditional Generative Adversarial Network (CGAN), as well as a mixture of parametric distributions, and we compare our models against previous work. 
Qualitatively and quantitatively, we show that the proposed approaches consistently outperform previous work, providing more realism and responsiveness. 

\end{abstract}

\maketitle

\section{Introduction}
Financial markets are among the most complex systems in existence. Naturally described as multi-agent systems, they comprise thousands of interacting heterogeneous participants.
%
Nowadays, both researchers and traders heavily rely on artificial market models, to support the design of algorithms, as well as testing novel trading strategies.
%
%
Artificial market models can help to isolate and study the impact of new algorithms to the price and volume of the stocks \cite{mizuta2020agent}; they can explain the nature of some rare financial market phenomena, such as bubbles and crashes \cite{paulin2018agent}; or they can just be used to study and test trading strategies, before approaching the real market \cite{coletta2021towards}.

Previous work mostly focuses on multi-agent modeling, which is a natural bottom-up approach to emulate financial markets \cite{farmer2009economy}. In these models, a number of decision-makers (agents or traders) and institutions, interact through prescribed rules to build the market.
%
Several multi-agent simulators have been developed, by traders and researchers \cite{byrd2020abides, raberto2001agent, chiarella2002simulation}. However, modeling a realistic market through a multi-agent simulation is still a major challenge \cite{mizuta2016brief, farmer2009economy}. 
In fact, specifying how the agents should behave and interact in the simulation is not obvious. While some agents can be modeled following a common sense or historical analysis \cite{farmer2005predictive, vyetrenko2019get}, in general market participants adopt unknown proprietary trading strategies. 
Moreover, public available historical data does not include attribution to the various market participants, which makes the calibration of the agents challenging. 

\begin{figure}[t]
\vspace{0.2in}
\begin{overpic}[width=0.9\columnwidth]{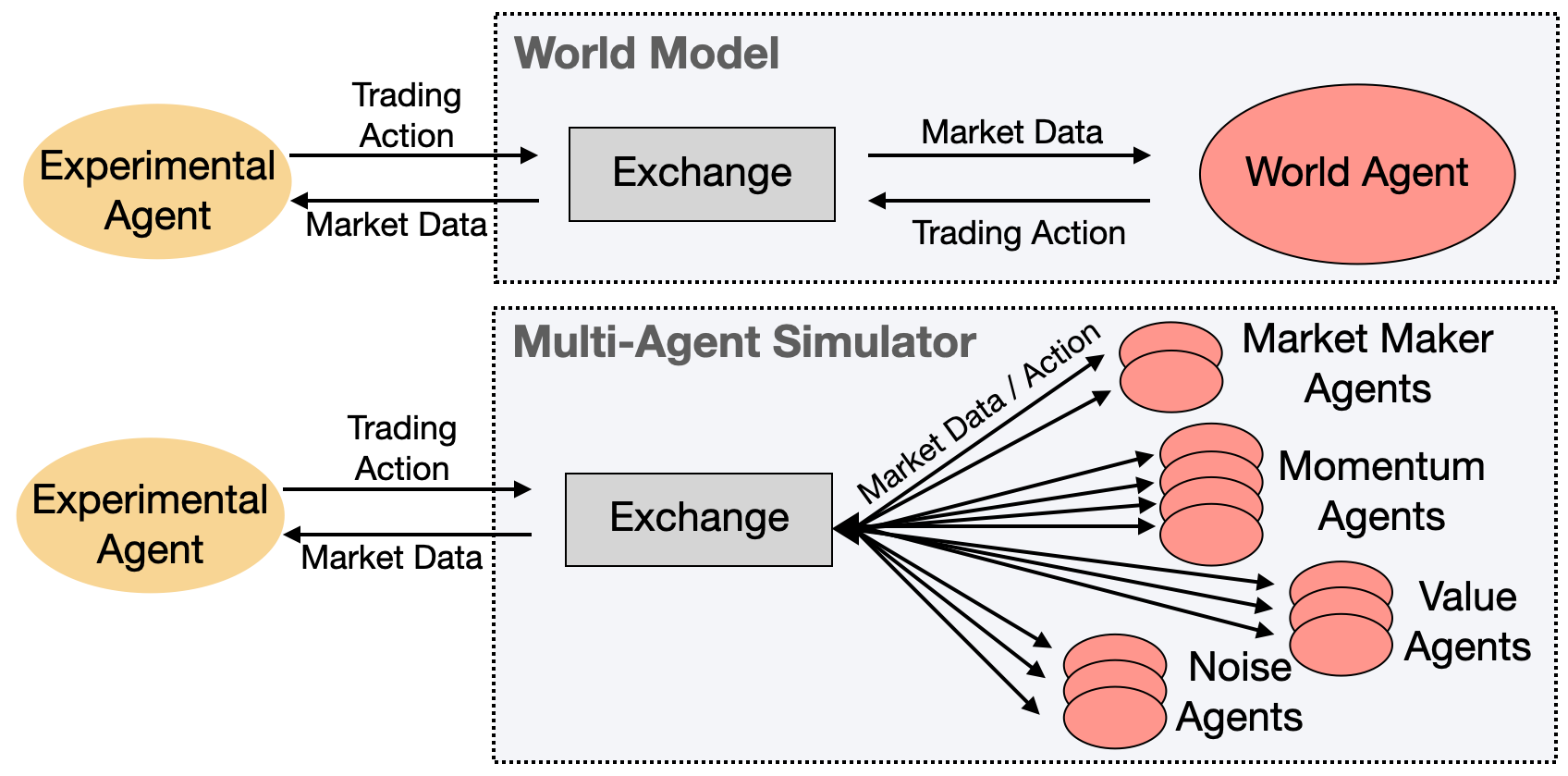}
 \put (0,45) {(A)}
 \put (0,22) {(B)}
\end{overpic}
\caption{World Model (A) vs Multi-Agent (B) Simulator.}
\vspace{-0.15in}
\label{fig:multi_vs_world_agent}
\end{figure}

To overcome this challenge, learning to simulate from the data as a \textit{world model} was introduced in \cite{coletta2021towards}. This approach assumed a unique agent trained as a conditional generative adversarial Network (CGAN) \cite{mirza2014conditional} from historical data, without the need of making assumptions about the individual market agent strategies.
\\
The world agent was simplistic: it was only capable of placing limit orders which usually account for just 50\% of all trading actions.
Nevertheless, it was shown in \cite{coletta2021towards} that this model could reproduce stylized facts as well as some form of market impact of trading - hence, this work provides a first attempt to a realistic and responsive \textit{world model}. 
Figure \ref{fig:multi_vs_world_agent} shows a classic multi-agent simulator compared to the \textit{World Model}, in which all the background agents are represented with a unique \textit{World Agent}.

In this paper, we improve the design of the CGAN-based world model by extending it to support all main market actions (i.e., market order, add limit order, cancel order, and replace order), and we describe it alongside another world model constructed explicitly as a mixture of parametric distributions. Moreover, we improve the CGAN robustness and stability by unrolling the model during the training. 

We demonstrate that both approaches presented in this paper outperform previous work on world model construction by providing higher degree of simulation realism. 
We also emphasize and experimentally demonstrate that the GAN-based model applies to different heterogeneous stocks, as it does not make explicit assumptions about the data distributions.

\section{Background}
In this section we introduce the readers to limit order book markets, with a brief introduction to market structure and mechanisms. 

\begin{figure}
 \centering
   \centering   \includegraphics[width=1\linewidth]{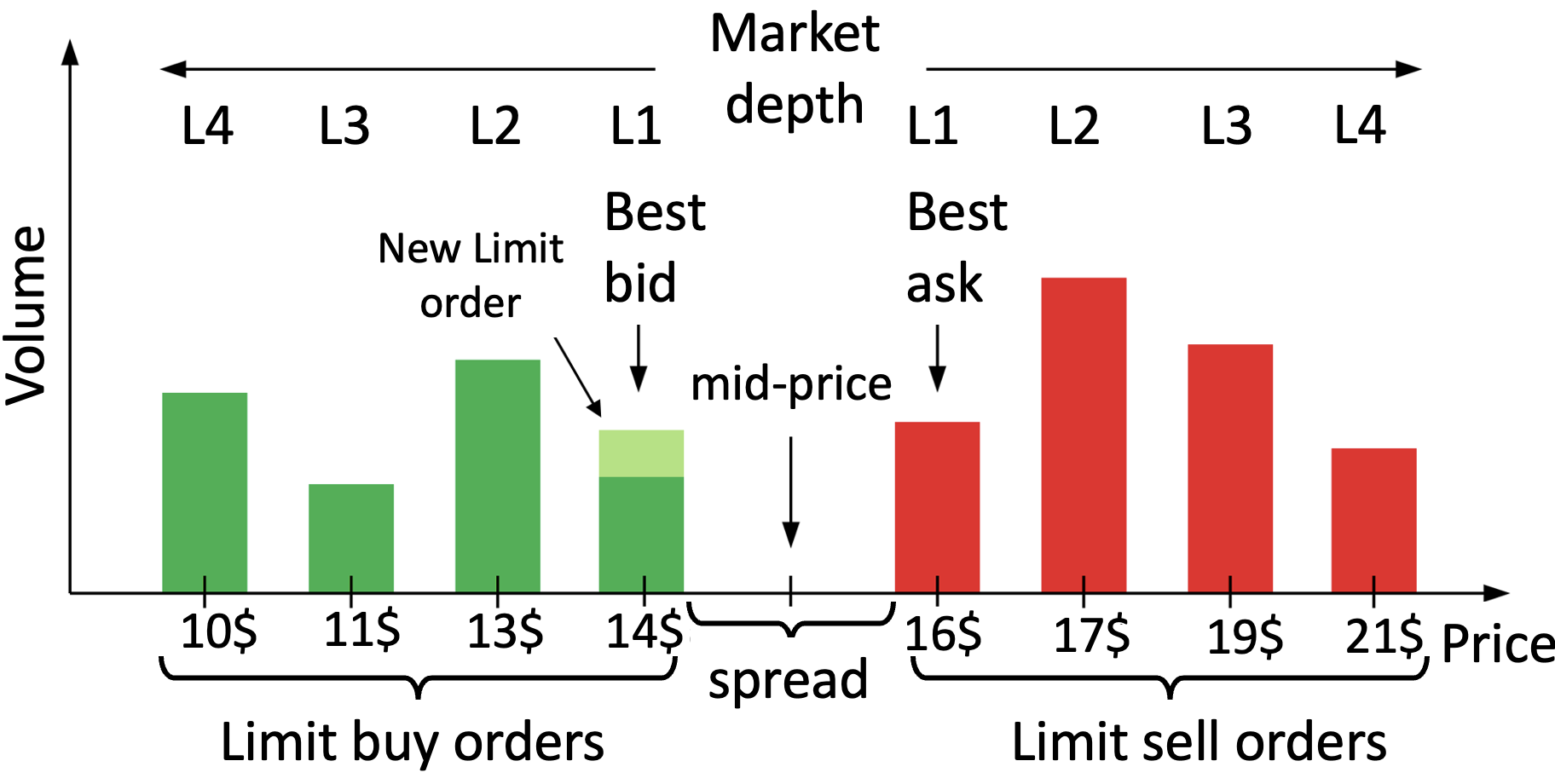}
   \caption{Order book definitions}
   \vspace{-0.1in}
 \label{fig:lob}
 \end{figure}
 
\subsection{Limit Order Book (LOB)}
Financial markets offer a place for buyers and sellers to meet and trade on different assets. 
Modern electronic markets, such as NASDAQ, provide ad-hoc message protocols to facilitate trades and provides real-time information about the market order flow and state. 
In particular, the ITCH \cite{nasdaq2020nasdaq} protocol provides access to anonymized market data with highest granularity, including all the orders in the market. 
The main four fundamental orders are: \textit{Market orders}, \textit{Add limit orders}, \textit{Cancel orders}, and \textit{Replace orders}. They respectively indicate the intention of trading a given amount of shares at any price; the intention of trading shares at a fixed limit price; a cancellation of a previous limit order; and a modification to a previous limit order (e.g., a change in the price or quantity).

Most equity markets employ a \textit{continuous-time double auction} mechanism to handle the stream of orders, and to execute a transaction whenever a buyer and seller agree on the price \cite{bouchaud2018trades}.
To store the supply and demand for each asset, the market exchange uses an electronic record called \textit{limit order book (LOB)}. The LOB keeps record of all outstanding limit orders into different levels, organized by price, and it continuously updates them according to incoming orders. 
Figure \ref{fig:lob} shows a snapshot of a LOB with the available supply (red bars) and demand (green bars). The first bars (L1) represent the first level, the second bars (L2) the second level, and so on. 
Each bar keeps the outstanding orders into a queue structure. An \textit{add limit order} to buy will update the existing demand, increasing the queue size (see Figure \ref{fig:lob} light green); while a \textit{cancel order} will decrease the queue size, and consequently reduce supply or demand.

\subsection{Artificial Market properties}

\paragraph{\textbf{Realism.}} Evaluating trading strategies against poorly calibrated market models can lead to poor and misleading conclusions, potentially causing severe loss when we employ these strategies on real markets. 
To assess the realism of artificial models, researchers commonly evaluate their ability to reproduce statistical properties of real markets called \textit{stylized facts} \cite{vyetrenko2019get, cont2001empirical, bouchaud2018trades}.
For example, as asset daily returns usually have fat tail distribution and long-range dependence, we expected the same properties (or \textit{stylized facts}) from artificial markets. In Section \ref{sec:eval} we show that our approach outperforms existing work under a wide range of stylized facts. In particular, we consider \textit{auto-correlations}, \textit{heavy tails distribution}, and \textit{long range dependence} to evaluate asset return properties. While we consider \textit{order volumes}, \textit{time to first fill}, \textit{depth} and \textit{market spread} distributions, to evaluate the volumes and order flow.  \footnote{We refer the reader to the work in \cite{vyetrenko2019get} for a detailed introduction to stylized facts.}

\paragraph{\textbf{Responsiveness.}} Another desirable property of artificial market models is the \textit{responsiveness} to exogenous trading orders: the model should emulate the market reaction to new orders, providing a tool to investigate strategies' impact on the market.
For example, the arrival of several buy (sell) market orders commonly causes the rise (fall) of the price. This phenomenon is called \textit{price impact}, and it desirable that a responsive model exhibit this behavior.

In section \ref{sec:eval} we evaluate the responsiveness of our model by simulating the arrival of a burst of buy/sell orders \cite{balch2019evaluate}.  

\subsection{Generative Models and CGANs}
In the last years generative models have been successfully employed in a wide range of scenarios, ranging from images to time-series. 
A generative model is any model able to learn a probability distribution $p_{\texttt{model}}$ resembling the real data distribution $p_{\texttt{data}}$, from a set of real samples. Among generative models, we can identify two major approaches: a) models that \textit{explicitly} estimate the probability density function; b) and models that  \textit{implicitly} learn to generate samples without the need of an explicit density function \cite{goodfellow2016nips}.

\paragraph{\textbf{Generative Adversarial Networks (GANs)}} GANs are powerful generative models that consider two adversarial neural networks,
which \textit{implicitly} learn to generate data samples \cite{NIPS2014_gan}.
In particular, a generator $G$ and a discriminator $D$ are trained simultaneously to compete in the following min-max game:
\begin{equation*}
   \min_G \max_D \, \underset{\textbf{x} \sim p_{\texttt{data}}((\textbf{x}))}{\mathbb{E}}[\log(D((\textbf{x})))] + \underset{\textbf{z} \sim p_{z}(\textbf{z})}{\mathbb{E}}[\log(1 - D(G(\textbf{z})))] 
\label{eq:gan_game}
\end{equation*}
The generator $G(z)$ produces new realistic samples from a prior noise distribution $p_{z}(\textbf{z})$, while the discriminator distinguishes between real and synthetic samples. 
Both networks aim at maximizing their own utility function: as the training advances, the discriminator $D$ learns to reject synthetic samples generated by $G$, which in turn learns to generate more realistic data to fool the discriminator. This two-player game trains the model to generate realistic samples resembling the real data. 

Our world model aims at generating realistic market actions in accordance with the current market, to provide realism and responsiveness to trading orders. Therefore, we consider a \textit{conditional} GAN \cite{mirza2014conditional}.
A CGAN generates realistic samples conditioned by some extra information $\textbf{y}$, representing the market state in our case. It will include information about ongoing events, including outstanding trading orders. Both generator and discriminator incorporate this extra information resulting into the following game:
\begin{equation*}
   \min_G \max_D \, \underset{\textbf{x} \sim p_{\texttt{data}}(\textbf{x})}{\mathbb{E}}[\log(D(\textbf{x}|\textbf{y}))] + \underset{\textbf{z} \sim p_{z}(\textbf{z})}{\mathbb{E}}[\log(1 - D(G(\textbf{z}|\textbf{y})))]
\label{eq:cond_gan_game}
\end{equation*}

\section{Related Work}

Artificial market models are increasingly adopted and developed to support researchers, practitioners, and traders \cite{farmer2009economy, byrd2020abides, coletta2021towards, mizuta2016brief, wang2017spoofing, wang2017stockyard, cho2021bit}.

Interactive Agent-Based Simulators (IABS) are popular models that represent the market by means of a set of trading agents that interact through prescribed rules. These simulators usually describe the agent population and their interaction by a set of common sense hand-crafted rules, made to mimic real markets \cite{vyetrenko2019get}. 
\\
In \cite{farmer2005predictive} Farmer et al. show how zero intelligence (ZI) agents are able to replicate some market dynamics, (i.e., spread and price) by placing random orders.
ZI agents have no knowledge of the market, however they can emulate some market reactions to exogenous events, and Palit et al. \cite{palit2012can} show that they also reproduce some stylized facts of real markets, like the fat tails and long-range dependencies.
Recently, M.P. Wellman and X. Wang \cite{wang2017spoofing} use ZI agents along heuristic belief learning (HBL) agents, which actually exploit order book information, to simulate and study spoofing attacks. 
\\
However, the recent work of S. Vyetrenko et al. \cite{vyetrenko2019get} discusses how the calibration of a realistic population of agents is a challenging task, as historical labeled data, with details about traders and their individual strategies, are not publicly available. The authors study different hand-crafted agent configurations, and they propose related stylized facts to evaluate the realism of simulation. In Section \ref{sec:eval} we evaluate our world model against classic multi-agent simulators, using an agent configuration proposed in \cite{vyetrenko2019get}.

Other work on market modeling leverages learning approaches, either by using adaptive agents, which learn from experience and evolve towards a more realistic behavior \cite{lo2005reconciling, lebaron2007long}, or by directly generating synthetic LOB data \cite{li2020generating, coletta2021towards}.

The work of J. Li et al. \cite{li2020generating} shows a first attempt to generate realistic LOB data. However, they focus on synthetic data generation, not considering market simulation and properties like responsiveness. 
The work of Coletta et al. \cite{coletta2021towards} introduces the \textit{world model} to learn how simulate a realistic market from data. This approach assumed a unique agent trained as a CGAN from historical data, without the need of individual market agent strategies or data. However, the model considers only add limit orders, which account for just 50\% of all trading actions, resulting in a partial market representation. We extend this work to all the main trading actions, and we evaluate it against our proposal in Section \ref{sec:eval}.

\section{The World Model}\label{sec:model}

The \textit{world model} provides a novel approach to market simulation: it considers a unique \textit{world agent} trained on historical data to emulate the whole traders' population, without the need of individual market agent strategies.
The \textit{world agent} observes the market state and generates the next trading action emulating the real traders' behavior. Figure \ref{fig:multi_vs_world_agent} shows the \textit{world agent} representing the whole ensemble of trading agents.

Formally, the world agent can be described as a conditional probability distribution $\mathcal{F}(\textbf{x} | \textbf{y})$ that generates the next market action $\textbf{x}$ given some information $\textbf{y}$ about the market. 
\\
The action $\textbf{x}$ represents a trading order to the exchange, which advances the simulation into a new market state. Thus, by iteratively generating new orders the simulation advances in time, exploiting the world agent to generate the market. 
 
\paragraph{\textbf{Actions}} We consider 4 possible actions representing the main trading orders, which are defined as follows:
\begin{itemize}
    \item \textbf{Add Limit Order} is a 3-tuple composed by <\textit{depth}, \textit{side}, \textit{quantity}>
    \item \textbf{Market Order} is a 2-tuple composed by <\textit{side}, \textit{quantity}>
    \item \textbf{Cancel Order } is a 3-tuple composed by <\textit{cancel depth}, \textit{side}, \textit{queue position}>  
    \item \textbf{Replace Order } is a 5-tuple composed by <\textit{cancel depth}, \textit{side}, \textit{queue position}, \textit{new depth}, \textit{new quantity}>
\end{itemize}

\noindent We introduce $p^i_a(t)$, $v^i_a(t)$, $p^i_b(t)$, $v^i_b(t)$ as the price and volume size at i-th level of the LOB, at time $t$, for ask and bid respectively. The \textit{depth} $d(t)$ of a limit order describes the order price $p(t)$ with respect to the best-bid and best-ask as follows:
\begin{equation}
  p(t) = \begin{cases}
    p^1_b(t) - d(t), & \text{If side = BID}\\
    p^1_a(t) + d(t), & \text{Else}
  \end{cases}
\label{eq:depth}
\end{equation}
we consider \textit{depths} rather than prices to improve model stability: depths are almost stationary, conversely to prices that change over time.
\\
The \textit{side} and \textit{quantity} describe the amount of shares and the side of an order (i.e., bid or ask), respectively. 
\\
\textit{Cancel depth} and \textit{queue position} are used to cancel and replace orders, as they accurately describe cancellation and replacement dynamics of real markets \cite{bouchaud2018trades}. In particular, the \textit{cancel depth} identifies the order book level, while the \textit{queue position} identifies the specific order at that level.

\paragraph{\textbf{Market state}} Modeling the market state plays the fundamental role of conditioning the generative model, to produce accurate and responsive actions. We introduce a set of features that best describe the current market $\textbf{s}_t$ at time $t$, and the ongoing events. 

We first define the \textit{volume imbalance} $I^i(t)$ as the demand and supply inequality within the first $i$ levels:
\begin{equation}
  I^i(t) = \frac{\sum_{j=1}^{i} v^j_b(t)}{\sum_{j=1}^{i} v^j_b(t) + v^j_a(t) }
\end{equation}\label{eq:vol_imbalance}
The volume inequality is a strong predictor of the future price change \cite{bouchaud2018trades}, and provides a view of the current market state.

Along the imbalance, we also define the  \textit{absolute volume} $V^i(t)$ within the first $i$ levels:
\begin{equation}
  V^i(t) = \sum_{j=1}^{i} v^j_b(t) + v^j_a(t)
\end{equation}\label{eq:abs_vol}
This feature helps the model balancing between cancel and limit orders, and generating accurate \textit{quantities} and \textit{depths}, to keep consistent volumes over the day. 

We define the \textit{order-sign imbalance} $O_{N}(t)$ for a history window of N events as follows:
\begin{equation}
  O_{N}(t) = \frac{1}{N} \sum_{j=t-N}^{t} \epsilon(j)
\end{equation}\label{eq:exec_imbalance}
where $\epsilon(j)$ is the sign of a market order at event-time $j$, if any. We consider $\epsilon(j) = 1$ for a sell market order, and $\epsilon(j) = -1$ for a buy market order. This feature provides knowledge about the price trend in the recent history, and about price impact phenomena.

We consider the market \textit{spread} $\delta(t)$, defined as: 
\begin{equation}\label{eq:spread}
  \delta(t) = p^{1}_a(t) - p^{1}_b(t)
\end{equation}
It helps the model balancing between liquidity provider and liquidity taker behavior, and to shape order \textit{depths}.
 \\
Finally, we consider the \textit{price return} $r_{N}(t)$ for a history window of $N$ events, defined as follows:
\begin{equation}
  r_{N}(t) = \frac{m_{(t)}}{m_{(t-N)}} - 1
\end{equation}\label{eq:return}
The returns describe the current market trends. 

\subsection{Explicit model: mixture of parametric distributions}\label{sec:expl_model}
\begin{figure}[t]
\includegraphics[width=1.05\linewidth]{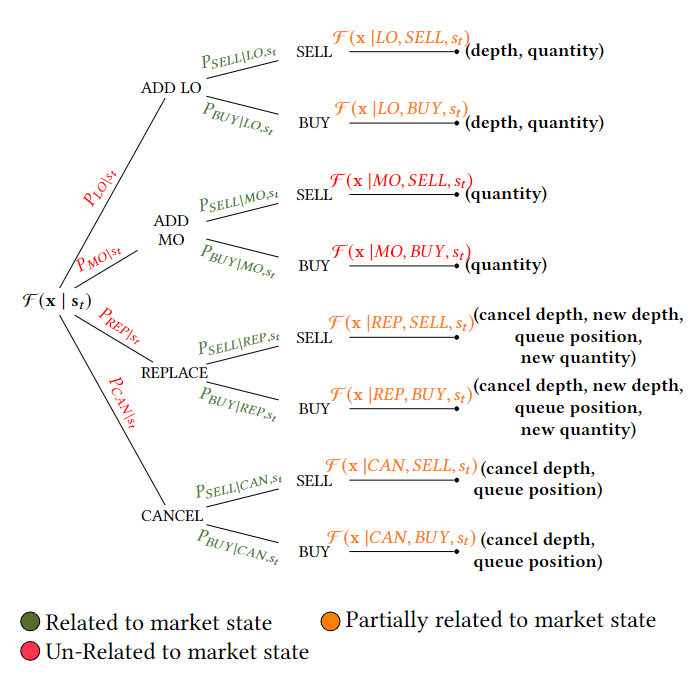}
\caption{Explicit model : mixture of parametric distributions.}
\label{fig:expl_model}
\vspace{-0.1in}
\end{figure}

We now introduce a simple and understandable world agent model based on classic parametric distributions.
We observe that the considered trading actions are composed of ordinal features with an unbounded range (e.g. quantity) but also relatively well-balanced categorical features (i.e., side and order type). Therefore, we consider a world agent expressed as a product of successive conditional distributions. We first condition the generation process with categorical features, which allows us to break down the order distributions into smaller conditional pieces, which are modeled with simple distributions. Figure \ref{fig:expl_model} shows the proposed approach in which the \textit{order type} and \textit{side} break down the complexity of the generation process, and condition the ordinal features (e.g., depth and quantity). Notice that, we use the following abbreviations: \textit{LO}, \textit{MO}, \textit{REP}, and \textit{CAN}, to identify add limit orders, market orders, replace orders, and cancel orders, respectively. 

The decomposition makes the world agent easier and more understandable: we can fit each distribution directly on the data, and independently from other distributions. We use classic and well-studied distributions, which have been carefully chosen after an accurate analysis of data, and according to existing literature \cite{bouchaud2018trades, vyetrenko2019get}. We use closed-form maximum likelihood or moment matching estimators to fit the distributions' parameters. 
For example, $P_{LO|s_t}$ is fitted as a fixed probability estimated using empirical proportions of actions in the historical data. At the second level of conditioning, $P_{SELL|LO,s_t}$ is fitted using only historical data in which the trading actions are limit order placements. To easily decompose the problem, and make it tractable and understandable, we make the following assumptions: - for limit order placement and replacement, we assume that \textit{depth} and \textit{quantity} are independent; - for order replacement, we assume that \textit{new depth} and \textit{new quantity} are independent from the former ones.
 
\begin{figure*}[t]
 \centering
   \centering
   \includegraphics[width=0.9\linewidth]{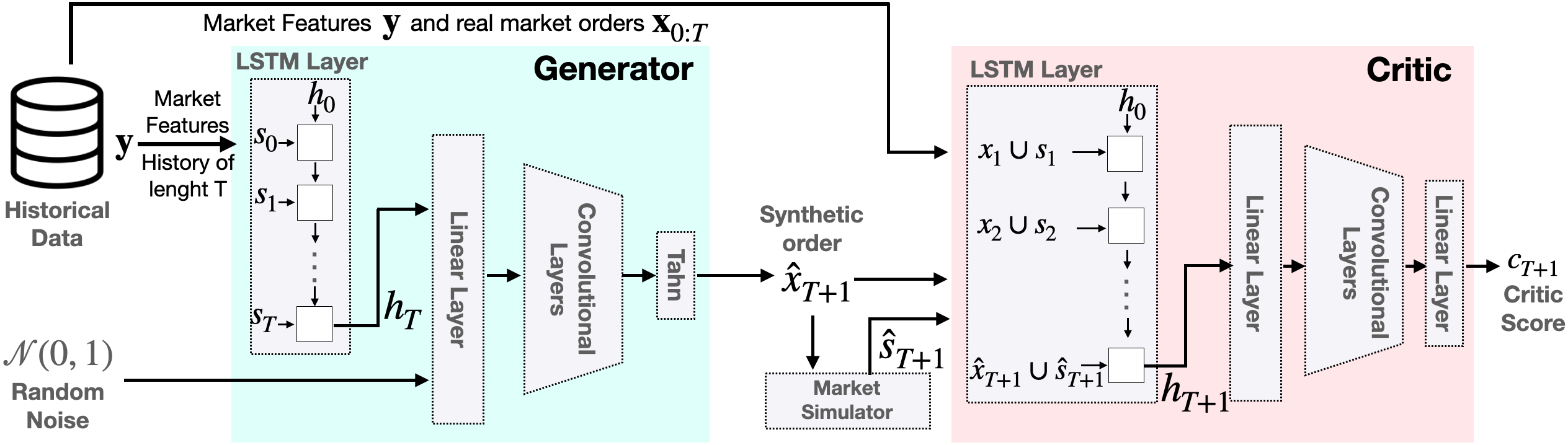}
   \caption{CGAN Architecture.}
 \label{fig:cgan_arch}
 \end{figure*}

We now introduce the distributions used for the categorical features. For the \textit{order type} we use a multinomial distribution fitted on historical data, while the \textit{side} consists of a binomial distribution conditioned on the order type and the volume imbalance $I^5(t)$, using a logistic model per each type. The logistic models show high probability of generating a BID limit order when $I^5(t) \approx 0$, which indicates that most of the volume is on the ASK side, and vice versa.
\\
Once we identify the order type and side, we use the following distributions to capture each specific order feature:
\begin{itemize}
\item  We describe the \textbf{depth} of a limit or replace order by a mixture of a beta-binomial distribution and an empirical multinomial distribution. The first distribution models negative depths while the latter accounts for positive depths. The probability of having a negative depth is modeled with a logistic regression, dependent on the market spread $\delta(t)$ and volume imbalance $I^5(t)$. 
\item We represent the order \textbf{quantity} as a mixture of two negative-binomial distributions. We observe that most of the investors trade quantities that are multiples of 100 shares, therefore one distribution describes quantities that are multiples of 100, and the other distribution is used for quantities that are not multiples of 100.
\item We represent the \textbf{cancel depth} using a negative-binomial distribution, considering replace and cancel orders separately. 
\item We model the \textbf{cancel queue position} using a beta-binomial distribution, considering replace and cancel orders separately.
\end{itemize}

\noindent Finally, we model the \textbf{inter-arrival time} of orders by fitting a gamma distribution on historical data. The inter-arrival time is the time between two consecutive orders, and it determines when the world agent is triggered to generate another trading action. Both our world models use this gamma distribution to model the inter-arrival times. In the experimental section, Figure \ref{fig:time_to_fill} shows the \textit{time to fill} for limit orders, which is strongly related to the order \textbf{inter-arrival times}. As both world models show a realistic time to fill, we can conclude that the gamma distribution generates \textbf{inter-arrival times} properly. 

\subsection{CGAN model}\label{sec:cgan_model}
We now introduce a CGAN-based \textit{world agent} implemented through a conditional Wasserstein GAN with gradient penalty (WGAN-GP) \cite{gulrajani2017improved}. 
We consider a WGAN-GP as it provides a more stable training and allows to deal with discrete data: it minimizes the Wasserstein-1 distance between real and synthetic data distributions, which is continuous and differentiable almost everywhere. 
In a WGAN-GP the discriminator does not classify samples, but it rather outputs a real value evaluating their realism, thus we refer to it as a \textit{Critic}. Note that in contrast with explicit model learning described in the previous section, the CGAN architecture does not take any parametric assumptions, and hence can be more easily extended to training on data that represents stocks with different dynamics.

\paragraph{\textbf{Model input.}} Our CGAN generator $G(\textbf{z}|\textbf{y})$ takes as input a vector of Gaussian random noise $z \sim \mathcal{N}(0,\,1)$ and a vector $\textbf{y}$ containing market information. We represent the market state at time $t$ as an ordered vector $\mathbf{s}_t$ defined as follows:
\begin{equation*}
    \mathbf{s}_t = \{I^1(t), I^5(t), O_{128}(t), O_{256}(t), V^1(t), V^5(t), \delta(t), r_{1}(t), r_{50}(t) \}
\end{equation*}\label{eq:gan_statea}
To capture the market evolution over time, we define  $\textbf{y}$ as the concatenation of the last $T$ historical market states: $\mathbf{y}_t = \{\textbf{s}_{t-T}, \ldots, \textbf{s}_{t}\}$. \\
Notice that, while most of the features take values in $[-1, 1]$, we normalize $V^1(t)$, $V^5(t)$ and $\delta(t)$ between -1 and 1, using a mix-max scaler.

\paragraph{\textbf{Model Output.}} The generator outputs a synthetic trading action, which may have different attributes upon its type: a \textit{market order} has a \textit{side} and a \textit{quantity}, while an \textit{add limit order} requires also a specific \textit{depth}. 
\\
To have an universal representation for all the orders, we consider an output vector $\mathbf{\hat{x}}$ including all the possible attributes:
\begin{equation*}
\mathbf{\hat{x}} = \text{\textit{(depth, cancel depth, qty$_{x}$, qty$_{100x}$, qty type, order type, side)}    }
\end{equation*}
The \textit{order type} assumes values in $\{-1,0,1\}$ and it discriminates between \textit{market orders}, \textit{add limit orders} and \textit{cancel orders}. In our CGAN architecture the \textit{replace orders} are represented and learned by a sequence of a cancel and an add limit order.     
The order \textit{quantity} is represented by 3-attributes, namely qty$_{x}$, qty$_{100x}$ and \textit{qty type}, and it is defined as follows:
\begin{equation}
  \text{\textit{quantity}} = \begin{cases}
    \text{qty$_{x}$}, & \text{If qty type = 1}\\
    100 \cdot \text{qty$_{100x}$}, & \text{Else}
  \end{cases}
\label{eq:quantity}
\end{equation}
As described in the previous section, most of the investors trade multiples of 100 shares, thus we use \textit{qty type} to discriminate their orders, and qty$_{100x}$ to learn the hundreds digits of the quantity.
The \textit{side} assumes values in $\{-1,1\}$ and it distinguishes SELL and BUY orders.
Finally, the \textit{depth} and \textit{cancel depth} assume discrete values in $\mathbb{Z}$, and they represent the price of the order to add or modify, respectively. To reduce the action space, the \textit{queue position} is predicted through a beta-binomial distribution considering its stable and regular distribution. 

Notice that, all the non categorical attributes are normalized between -1 and 1. Moreover, depending on the \textit{order type}, only some attributes are meaningful and used: for an \textit{add limit order} we consider both \textit{depth}, \textit{side} and \textit{quantity} attributes, but for a \textit{cancel order} we consider just the \textit{cancel depth} and the \textit{side}.

\paragraph{\textbf{Model training and architecture.}} Figure \ref{fig:cgan_arch} shows the proposed CGAN architecture, which extends the previous model presented in \cite{coletta2021towards}. In particular, our CGAN improves stability and responsiveness by unrolling the model during the training. 
In \cite{coletta2021towards} the model is trained using only ground truth market states: at each training iteration the CGAN receives a real market state $\textbf{s}_t$ to generate the next order $\textbf{x}_{t+1}$. At test time, when the model is employed, and unrolled in a closed-loop simulation, the CGAN may encounter unseen states induced by a previous sequence of sub-optimal actions. Unseen states can lead to poor and misleading simulation (e.g., exponential market growth).
To mitigate unseen and unrealistic market states, we feed the generated orders into a simulator that advances the market state, and we let the \textit{critic} evaluates both generated orders and states during the training. Most important, we unroll the model during training: we generate $k$ steps ahead, feeding the model with the previous generated market states. This approach enforces the model to deal with synthetic market states, improving stability and realism: actions that led to unrealistic states will be penalized, while we also minimize unseen states. Notice that, we increase the value of $k$ during the training epochs, as the generator learns to generate more realistic orders. 

\section{Experiments}\label{sec:eval}
In this section we evaluate our world model by comparing the two proposed approaches in terms of realism and responsiveness. 
We train our models using NASDAQ TotalView data \cite{totalview} sent via ITCH
protocol \cite{nasdaq2020nasdaq} replayed at a simulated exchange at the trading action level. 
We consider four small-tick stocks, i.e., \textit{AVXL}, \textit{AINV}, \textit{CNR} and \textit{AMZN}, we use 3 to 4 days of data to train the models, and 9 days for testing. The results are averaged for each stock, over the 9 days period. We implement our models extending ABIDES simulator \cite{byrd2020abides}, and we feed real data market from \textit{09:30} to \textit{10:00} to initialize the simulated market, and condition the models. For simplicity, we refer to the model implemented through a mixture of parametric distributions as the \textit{explicit model} (see Section \ref{sec:expl_model}), while we refer to the CGAN-based model as the \textit{CGAN model} (see Section \ref{sec:cgan_model}).

  \begin{figure}[t]
\centering
\centering
  \begin{subfigure}{0.445\linewidth} 
  \includegraphics[width=\textwidth]{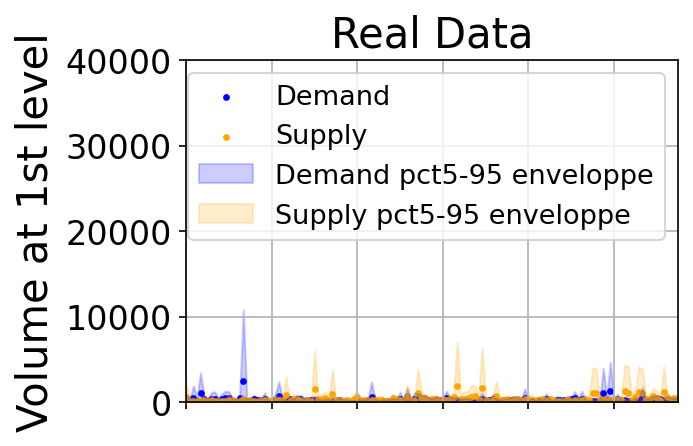} 
  \end{subfigure}  
  \centering
  \begin{subfigure}{0.33\linewidth}
  \includegraphics[width=\textwidth]{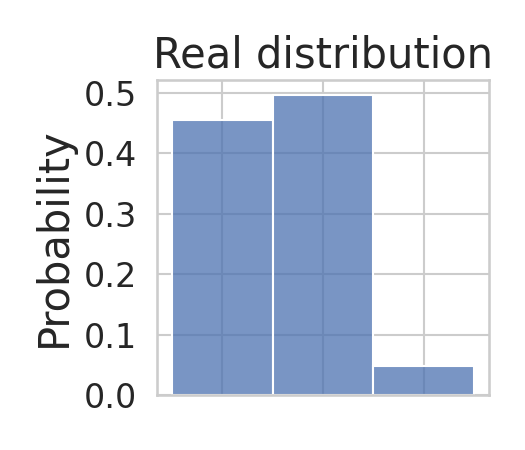}
  \end{subfigure}
  
\centering
  \begin{subfigure}{0.445\linewidth} 
  \includegraphics[width=\textwidth]{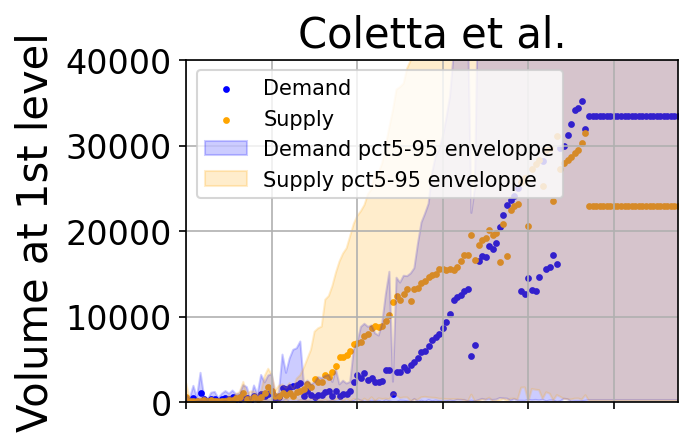} 
  \end{subfigure}  
  \centering
  \begin{subfigure}{0.33\linewidth}
  \includegraphics[width=\textwidth]{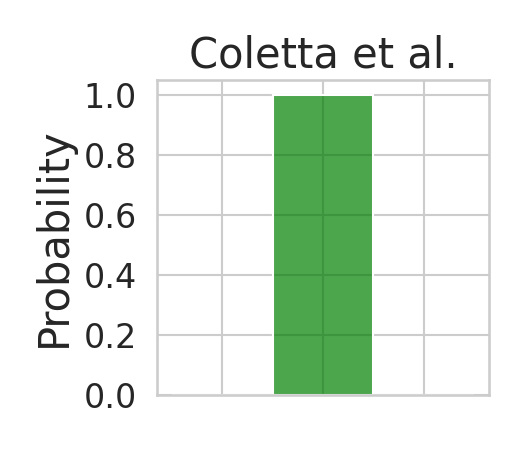}
  \end{subfigure}
  
    \begin{subfigure}{0.43\linewidth} 
  \includegraphics[width=\textwidth]{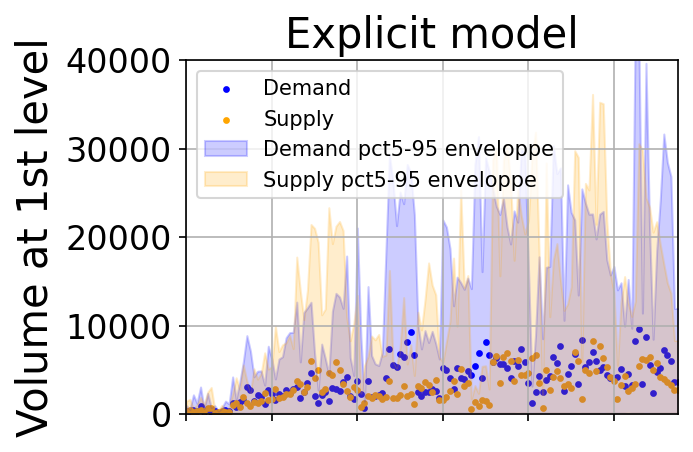} 
  \vspace{-0.15in}
  \end{subfigure}
\centering
  \begin{subfigure}{0.33\linewidth}
  \includegraphics[width=\textwidth]{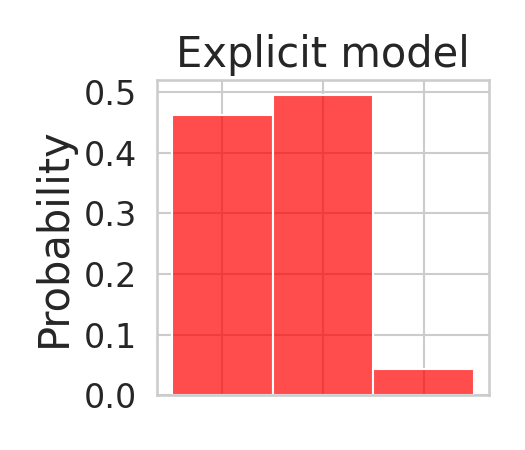}
\vspace{-0.15in}
  \end{subfigure} 

\centering
  \begin{subfigure}{0.445\linewidth} 
  \includegraphics[width=\textwidth]{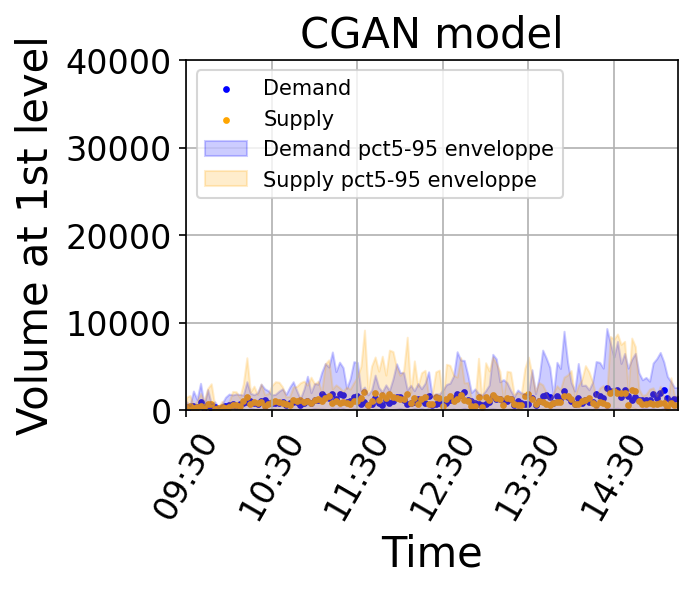} 
  \end{subfigure}  
  \centering
  \begin{subfigure}{0.33\linewidth}
  \includegraphics[width=\textwidth]{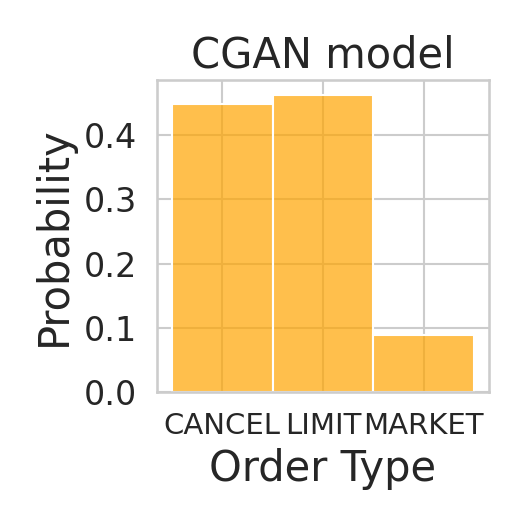}
  \end{subfigure}
  
  \caption{\small{Real vs Synthetic order distributions (AVXL). Proposed world models generate more realistic markets compared to previous work: they generate all main market actions which leads to more realistic volumes. }}\label{fig:coletta_comparison}
  \vspace{-0.2in}
  \end{figure}

\paragraph{\textbf{Previous work comparison}} We first compare our models against the previous work of Coletta et al. \cite{coletta2021towards}.  
Figure \ref{fig:coletta_comparison} \textit{right column} charts show the order types in the real and synthetic markets. While both our models closely resemble the market structure, the previous work of Coletta et al. \cite{coletta2021towards} (\textit{green chart}) only represents limit orders, accounting for just 50\% of all orders and lacking of realism. \footnote{We represent replace orders as a sequence of a cancel and a limit order to keep the representation consistent with CGAN model.} The \textit{left column} charts show the demand and supply (i.e., outstanding limit orders) at the first level of the order book, for real and simulated markets. 
The orange dots represent the average ASK volumes (supply), while the blue dots represent the average BID volumes (demand). The filled area represents the values between the 5th and 95th percentile. 
The charts clearly show unrealistic volumes that exponential increase for the work in \cite{coletta2021towards} (\textit{second chart}), mainly caused by the absence of cancel and market orders. Instead, our CGAN model faithfully reproduces real market volumes (\textit{fourth chart}), while the explicit model overestimates the volumes (\textit{third chart}) but it keeps reasonable average values (dots) and does not show exponential growth.  

 \begin{figure}[b]
\centering
  \begin{subfigure}{0.37\linewidth}
  \includegraphics[width=\linewidth]{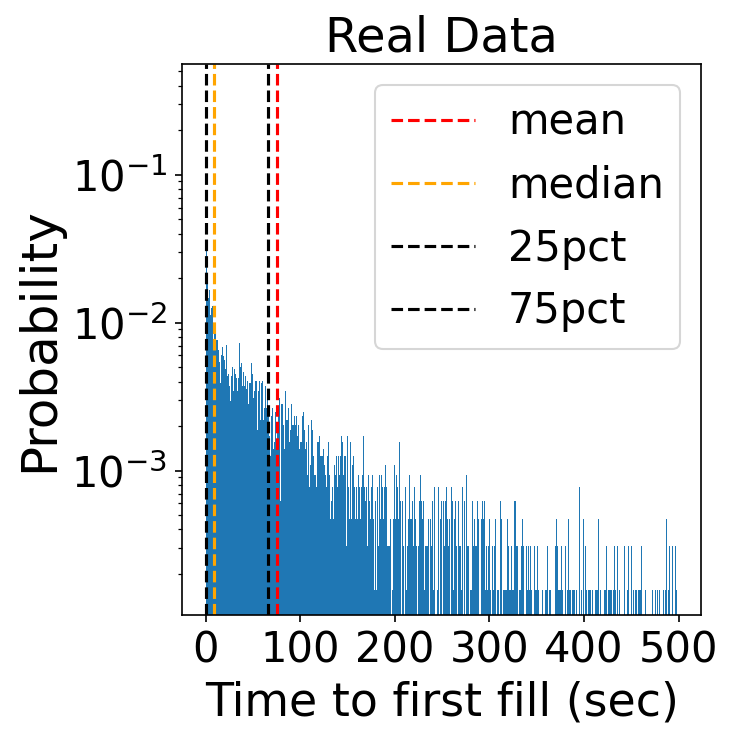}
  \end{subfigure}
  \hfill
\centering
  \begin{subfigure}{0.30\linewidth}
  \includegraphics[width=\linewidth]{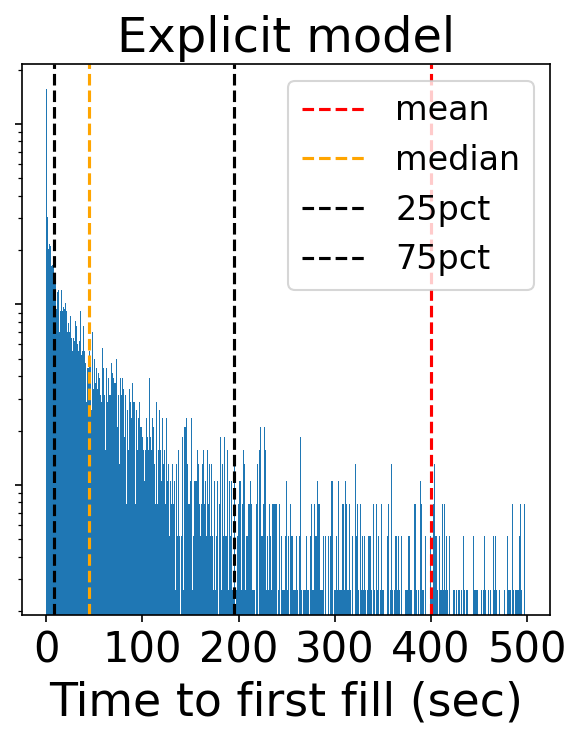}
  \end{subfigure} 
  \hfill
\centering
  \begin{subfigure}{0.30\linewidth} 
  \includegraphics[width=\linewidth]{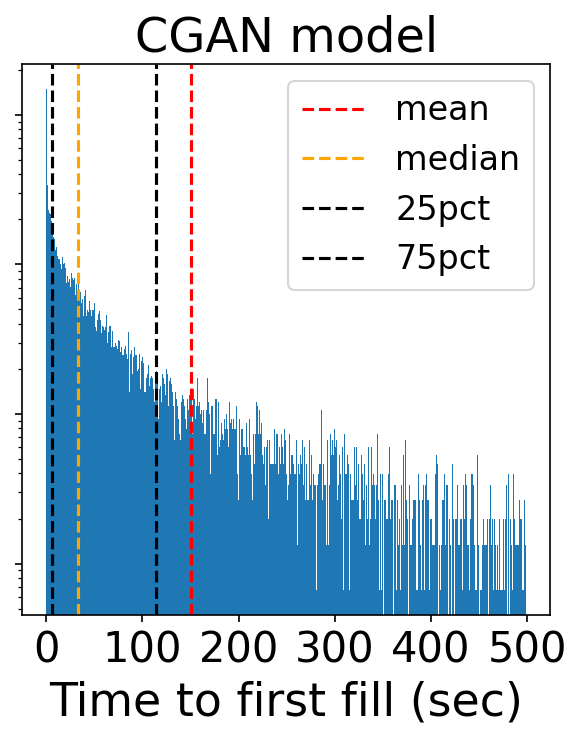} 
  \end{subfigure}  
      \vspace{-0.1in}
  \caption{\small{Time to first fill (seconds): CGAN model times are closer to historical ones.}}
  \label{fig:time_to_fill}
    \vspace{-0.1in}
\end{figure}

\paragraph{\textbf{Volumes and order flow stylized facts}} 
We now evaluate the ability of the explicit and CGAN model to reproduce a set of market stylized facts for a given stock, namely \textit{AVXL}.
We first investigate the \textit{time to first fill}, defined as the time elapsed between the placing of a limit order and its actual execution. This stylized fact describes how closely the synthetic market captures the real market dynamics and liquidity. Liquid markets usually have low \textit{time to first fill}, compared to less liquid ones. 
\\
Figure \ref{fig:time_to_fill} shows that both proposed models generate a realistic \textit{time to first fill}, with a median time less than 50 seconds (yellow vertical line). The CGAN model produces a slightly more accurate market compared to the explicit model: the average (red line) and 75th percentile (black line) values are closer to the real ones.   

\begin{figure}[h]
\centering
  \begin{subfigure}{0.37\linewidth}
  \includegraphics[width=\textwidth]{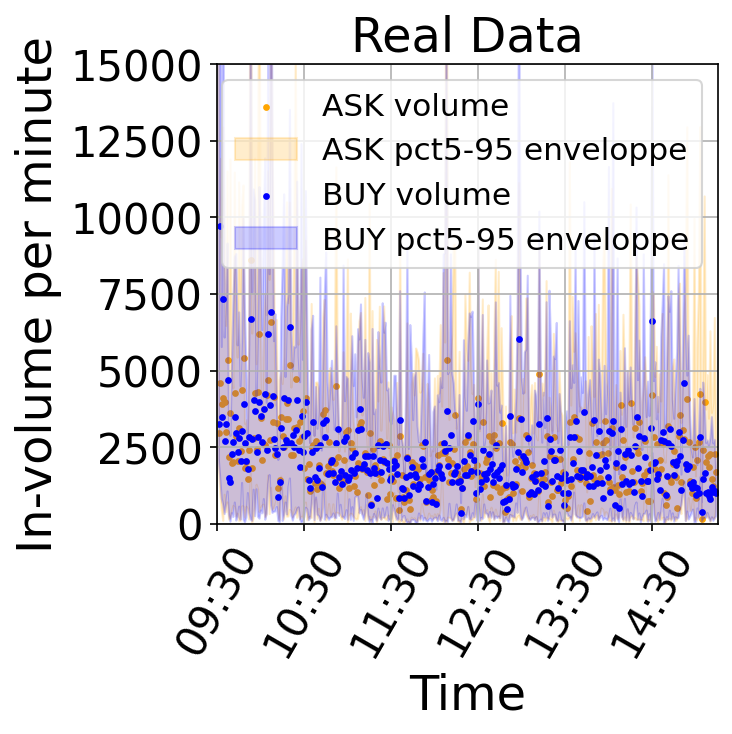}
  \end{subfigure}
  \hfill
\centering
  \begin{subfigure}{0.30\linewidth}
  \includegraphics[width=\textwidth]{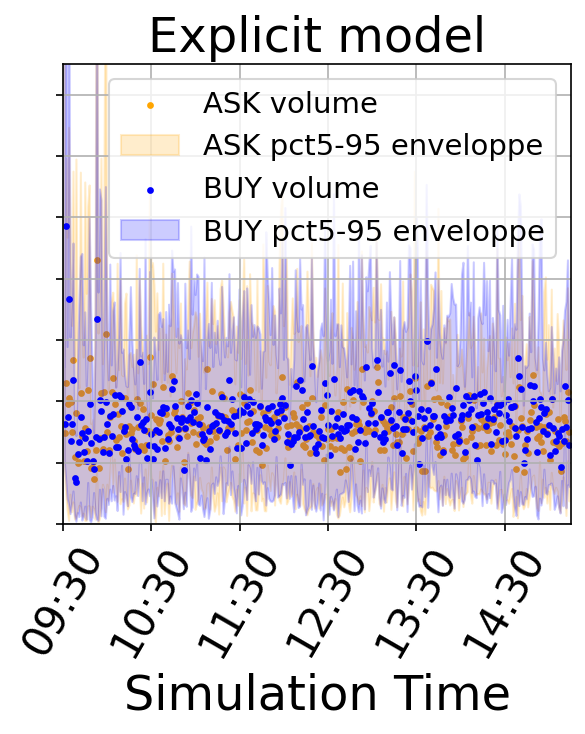}
  \end{subfigure} 
  \hfill
\centering
  \begin{subfigure}{0.30\linewidth} 
  \includegraphics[width=\textwidth]{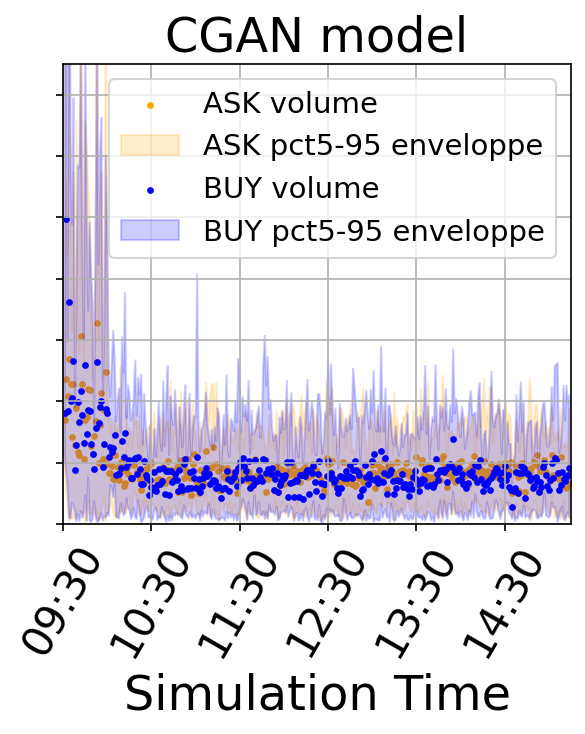} 
  \end{subfigure}  
    \vspace{-0.1in}
  \caption{\small{Aggregated volume of limit orders: both world models generate realistic volumes. }}
  \label{fig:mean_vol_lo}
    \vspace{-0.1in}
\end{figure}

Then, we analyse the incoming volume in real and simulated markets. Figure \ref{fig:mean_vol_lo} shows the volumes of add limit orders, aggregated in a minute time window: the dots represent the average incoming volumes per minute, while the filled area represents the 5th and 95th percentile values. 
This stylized fact represents the liquidity provided by market participants during the day. While both models closely resemble the real market, the explicit model slightly overestimates the volumes (as shown also in Figure \ref{fig:coletta_comparison}).

 \begin{figure}[h]
\centering
  \begin{subfigure}{0.36\linewidth}
  \includegraphics[width=\textwidth]{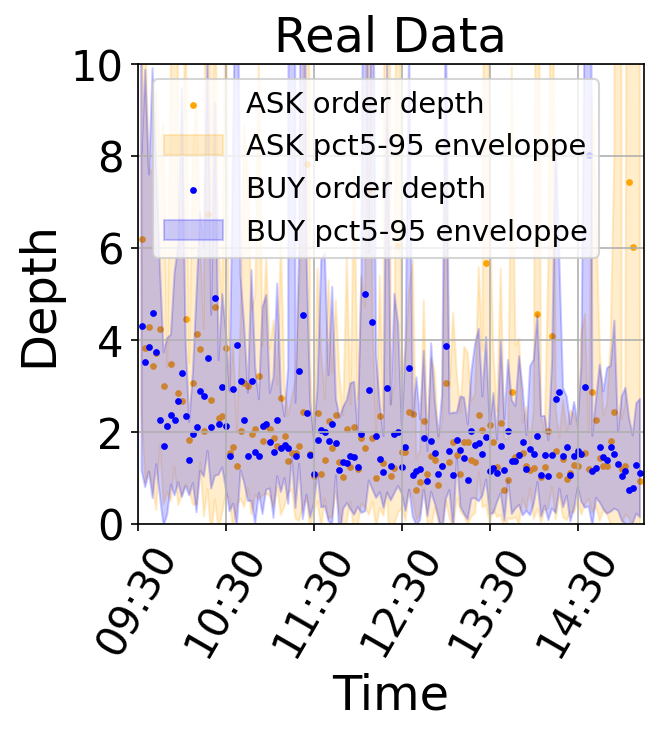}
  \end{subfigure}
  \hfill
\centering
  \begin{subfigure}{0.31\linewidth}
  \includegraphics[width=\textwidth]{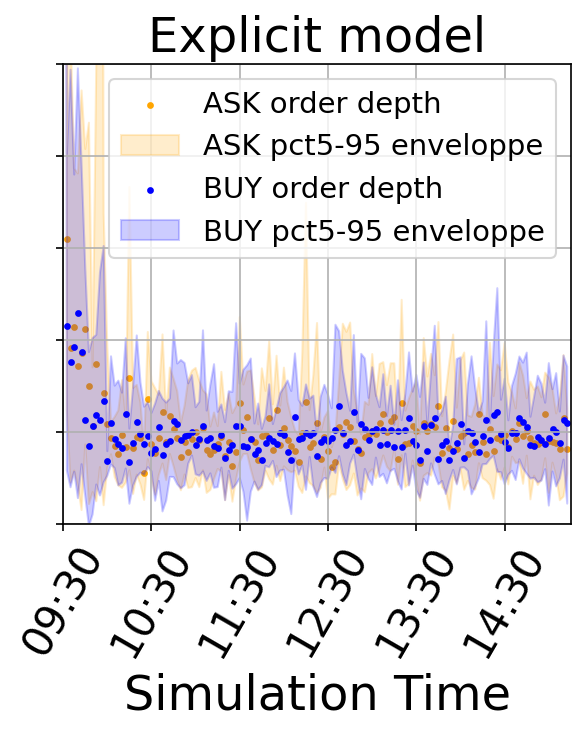}
  \end{subfigure} 
  \hfill
\centering
  \begin{subfigure}{0.31\linewidth} 
  \includegraphics[width=\textwidth]{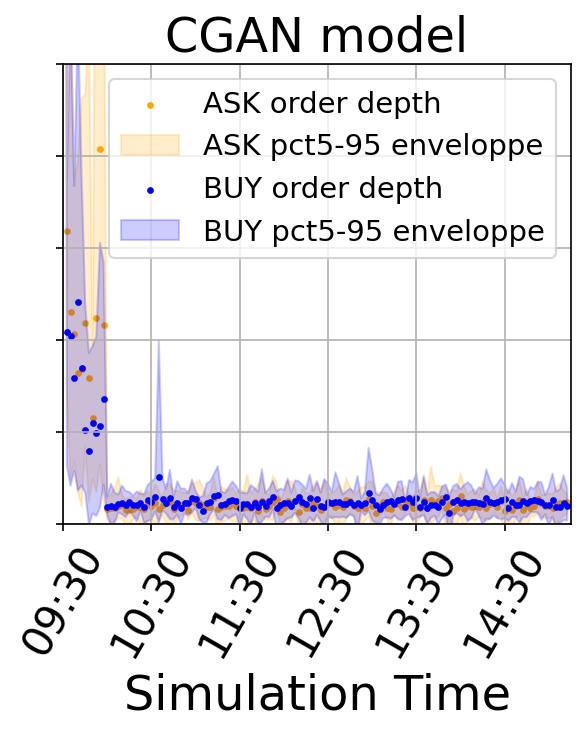} 
  \end{subfigure}  
    \vspace{-0.1in}
  \caption{\small{Depth of limit orders: explicit model  reproduces realistic order depths.}}
  \label{fig:depth}
\end{figure}
Figure \ref{fig:depth} shows the average depth of limit orders, for BID orders (blue dots) and ASK orders (orange dots). The chart shows how the explicit model closely resembles the real market data, even if the data have slightly less variance. Instead, the CGAN model only partially matches the real data: it shows a narrow data distribution, with most of the depths close to zero.

 \begin{figure}[h]
\centering
  \begin{subfigure}{0.36\linewidth}
  \includegraphics[width=\textwidth]{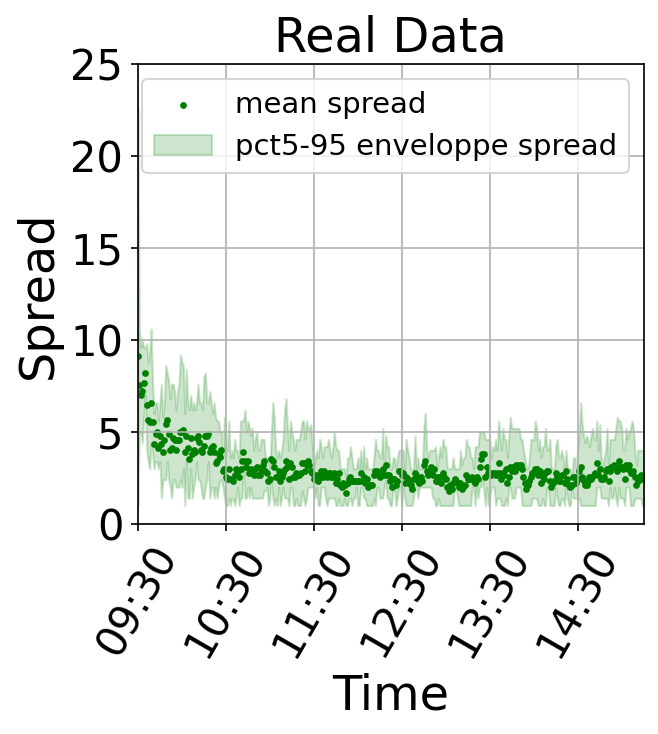}
  \end{subfigure}
  \hfill
\centering
  \begin{subfigure}{0.31\linewidth}
  \includegraphics[width=\textwidth]{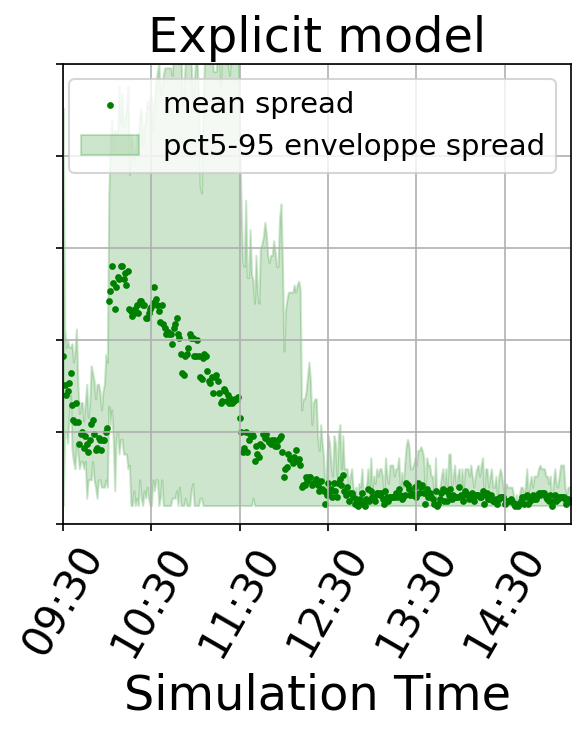}
  \end{subfigure} 
  \hfill
\centering
  \begin{subfigure}{0.31\linewidth} 
  \includegraphics[width=\textwidth]{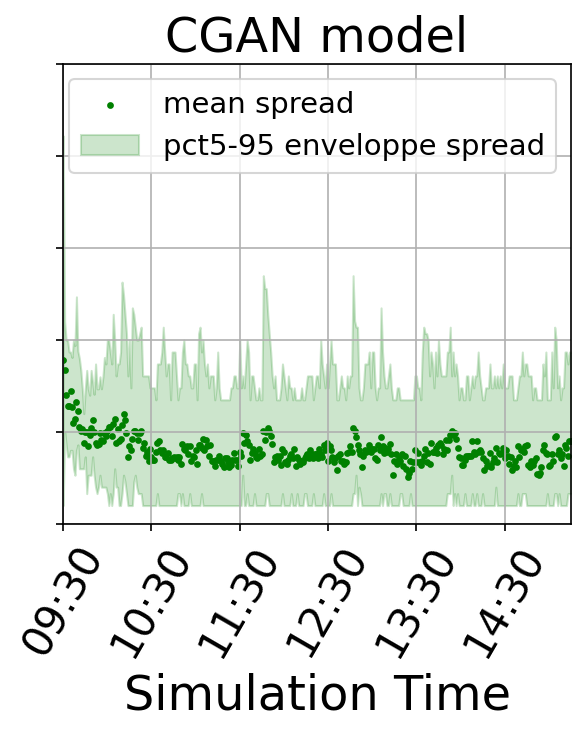} 
  \end{subfigure}  
    \vspace{-0.1in}
  \caption{\small{Market spread: CGAN model closely reproduces real market spread.}}
  \label{fig:spread}
  \vspace{-0.1in}
\end{figure}
Figure \ref{fig:spread} shows the market spread (see Eq. \ref{eq:spread}) over the day. The green dots represent the average spread in the real and simulated markets, and the filled area shows the 5th and 95th percentile values. The CGAN model has the best performance, it closely replicates the real market behavior, while the explicit model shows a slight adaptation problem to real market data. When first employed in the market at \textit{10:00} (after the market is initialized with real data), the explicit model doubles the market spread, which decreases and stabilizes only after \textit{12:00}. 

\begin{figure}[h]
\centering
  \begin{subfigure}{0.315\linewidth}
  \includegraphics[width=\textwidth]{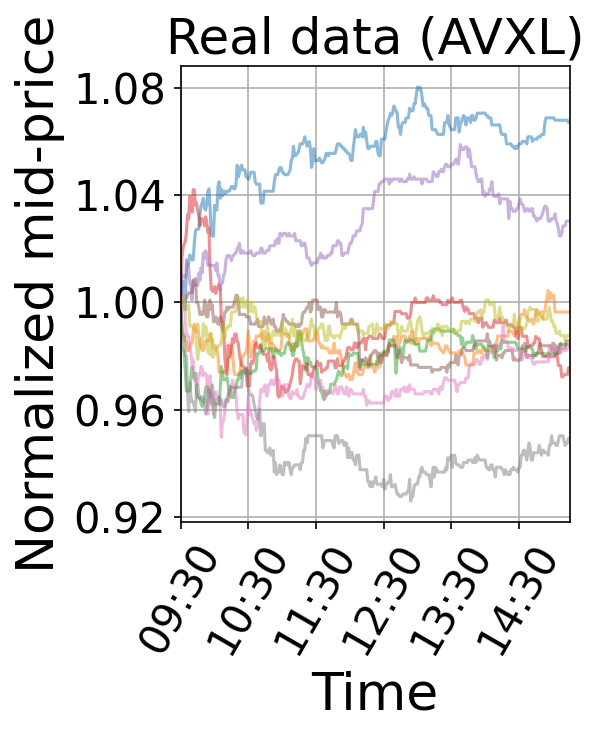}
  \end{subfigure}
  \hfill
\centering
  \begin{subfigure}{0.33\linewidth} 
  \includegraphics[width=\textwidth]{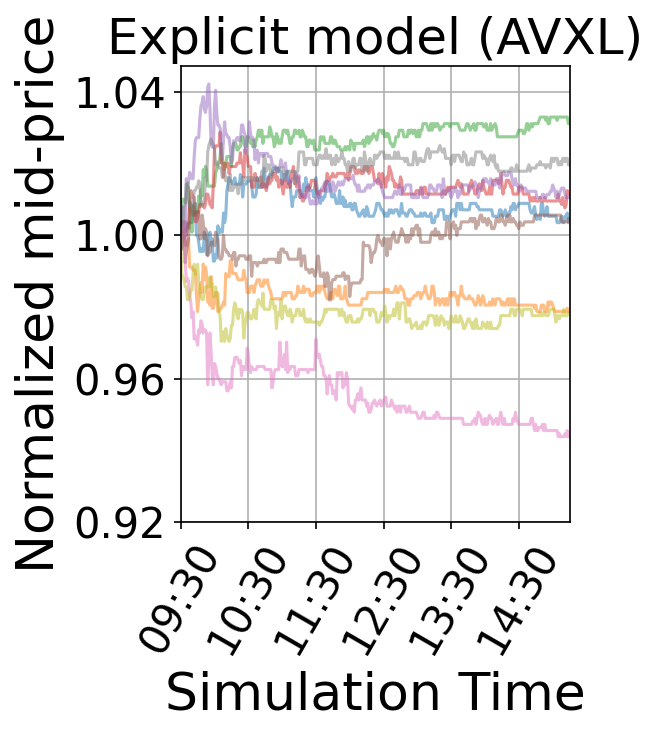} 
  \end{subfigure}  
    \hfill
\centering
  \begin{subfigure}{0.32\linewidth}
  \includegraphics[width=\textwidth]{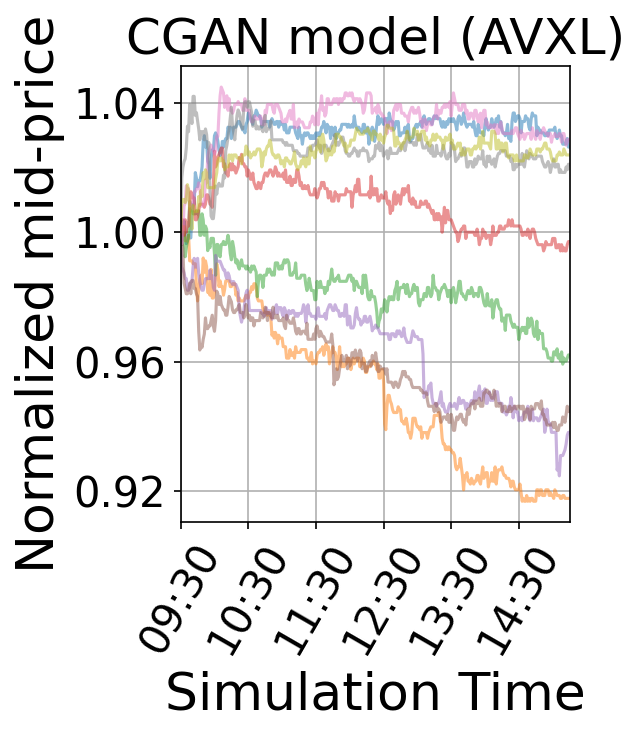}
  \end{subfigure} 
  \vspace{-0.1in}
  \caption{\small{Generated time-series (AVXL) show diversity and realism.}}
  \vspace{-0.1in}
\label{fig:timeseries}
\end{figure}

Finally, we show an example of the generated time-series in Figure \ref{fig:timeseries}. The charts show the normalized mid-price: the left chart shows the real samples, the middle chart shows the explicit model mid-prices, and the right chart shows the CGAN model ones. Both models shows promising diverse and realistic time-series data.

\begin{figure}[h]
\centering
  \begin{subfigure}{0.325\linewidth}
  \includegraphics[width=1.05\textwidth]{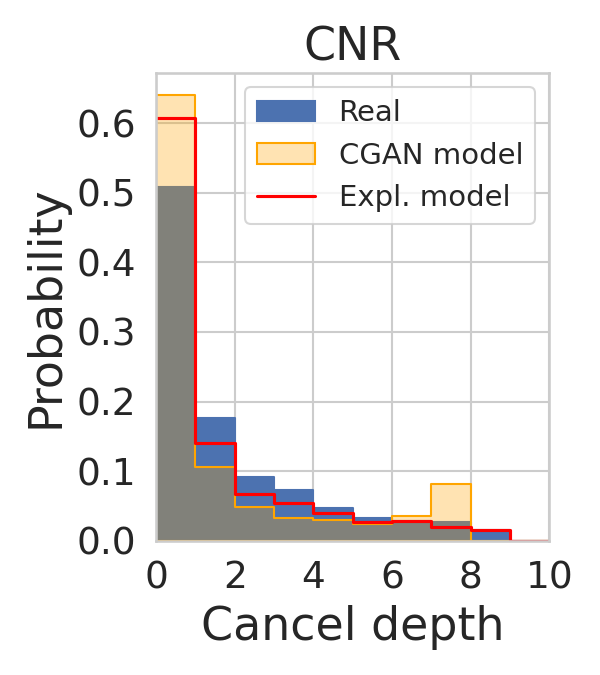}
  \end{subfigure}
  \hfill
\centering
  \begin{subfigure}{0.325\linewidth}
  \includegraphics[width=1.05\textwidth]{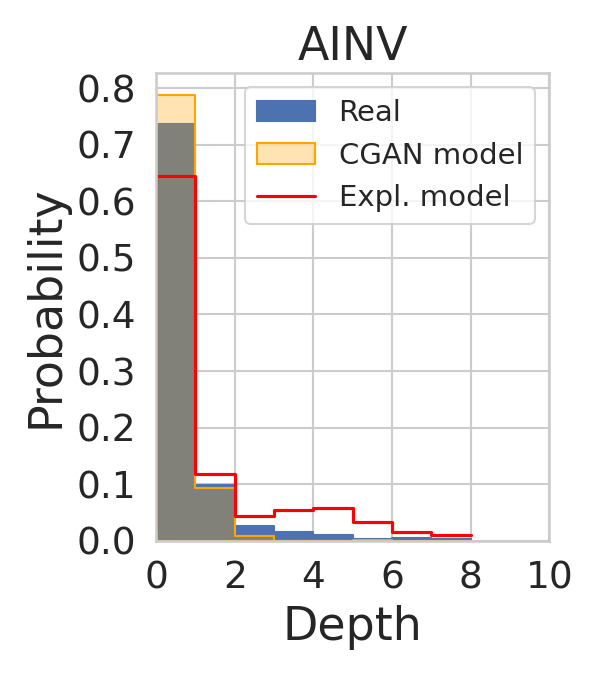}
  \end{subfigure} 
  \hfill
\centering
  \begin{subfigure}{0.325\linewidth} 
  \includegraphics[width=1.05\textwidth]{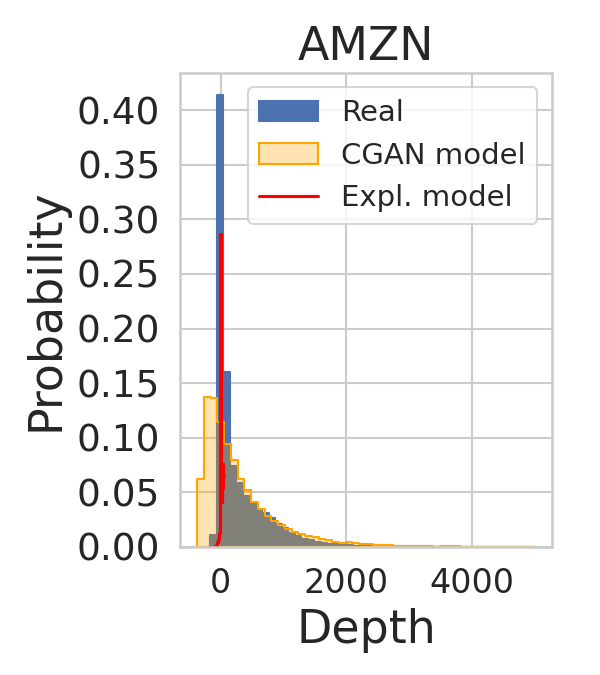} 
  \end{subfigure}  
\caption{\small{Training on different stocks (AINV, CNR, AMZN): the CGAN model adapts better compared to the explicit model.}}\label{fig:scalability}
\end{figure}

\paragraph{\textbf{Training on different stocks}} We now discuss how our models apply to different stocks. Figure \ref{fig:scalability} shows the orders generated by the explicit and CGAN model for three different stocks, namely CNR, AINV and AMZN. The blue bars show the real data distributions, the orange bars show the CGAN synthetic data, and the bars with red lines outline the explicit model synthetic data (we use empty bars to improve readability). 
The first two charts show the \textit{cancel depth} and \textit{depth} of CNR and AINV orders, respectively. The charts show that both proposed models are able to generate realistic data (i.e., the bars mostly overlap). 
The last chart shows the \textit{depth} distribution for AMZN orders. While the CGAN model is able to generate realistic data, the explicit model fails to reproduce the \textit{depth} of the orders (i.e., it generates only values close to 0). This chart shows the main advantage of the CGAN model: with fewer assumptions on underlying data structure, the model is able to represent a wider range of stocks, with different behaviors and distributions.   

  \begin{figure}[t]
\centering
  \begin{subfigure}{0.22\textwidth} 
  \begin{overpic}[width=\textwidth]{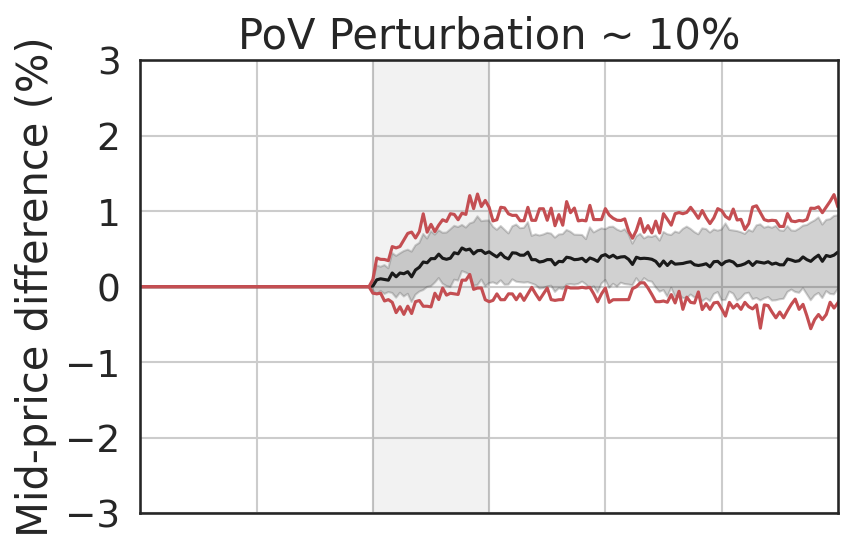} 
 \put (35,65) {Explicit Model}
\end{overpic}
  \end{subfigure}
  \hfill
\centering
    \begin{subfigure}{0.22\textwidth}
\begin{overpic}[width=\textwidth]{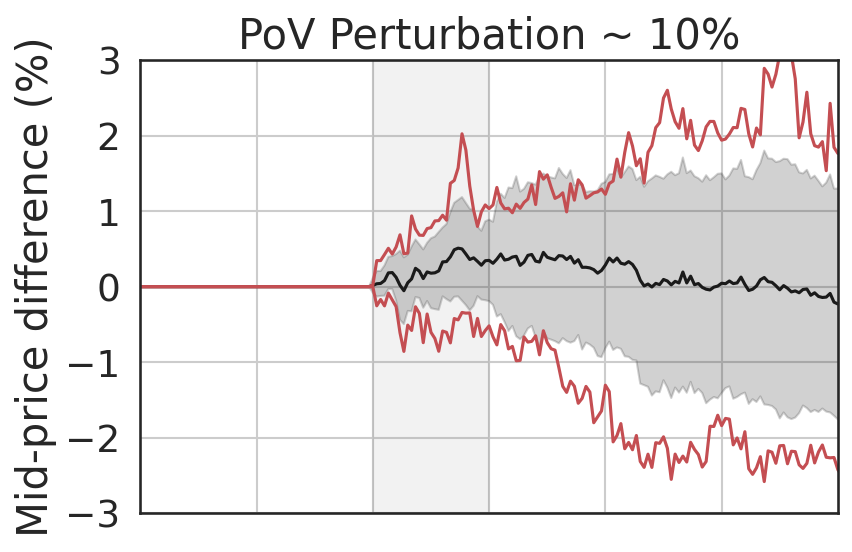} 
 \put (35,65) {CGAN Model}
\end{overpic}
  \end{subfigure} 
  \hfill
\centering
  \begin{subfigure}{0.22\textwidth} 
  \includegraphics[width=\textwidth]{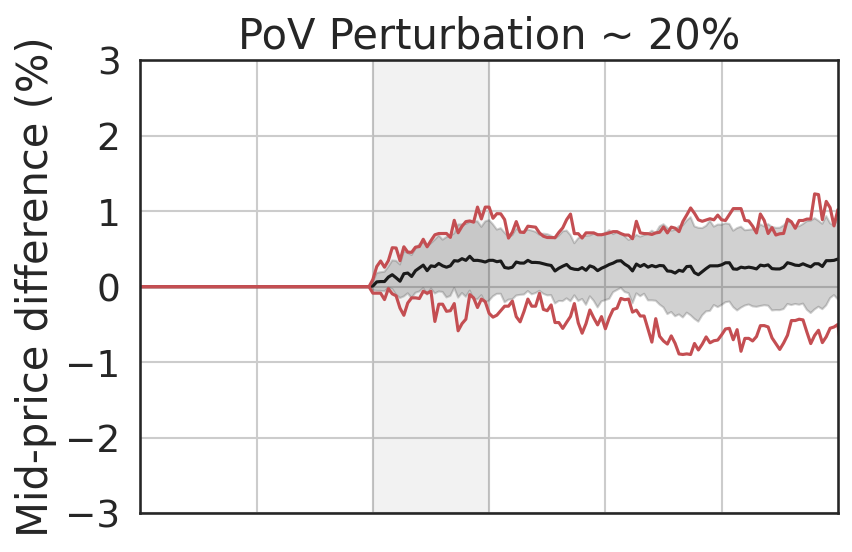} 
  \end{subfigure}  
  \hfill
  \centering
  \begin{subfigure}{0.22\textwidth}
  \includegraphics[width=\textwidth]{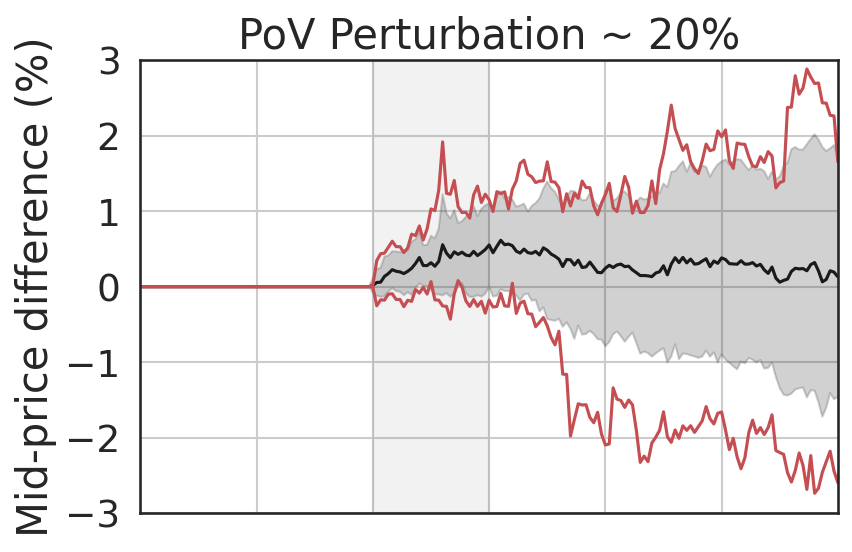}
  \end{subfigure}
  \hfill
\centering
  \begin{subfigure}{0.22\textwidth} 
  \includegraphics[width=\textwidth]{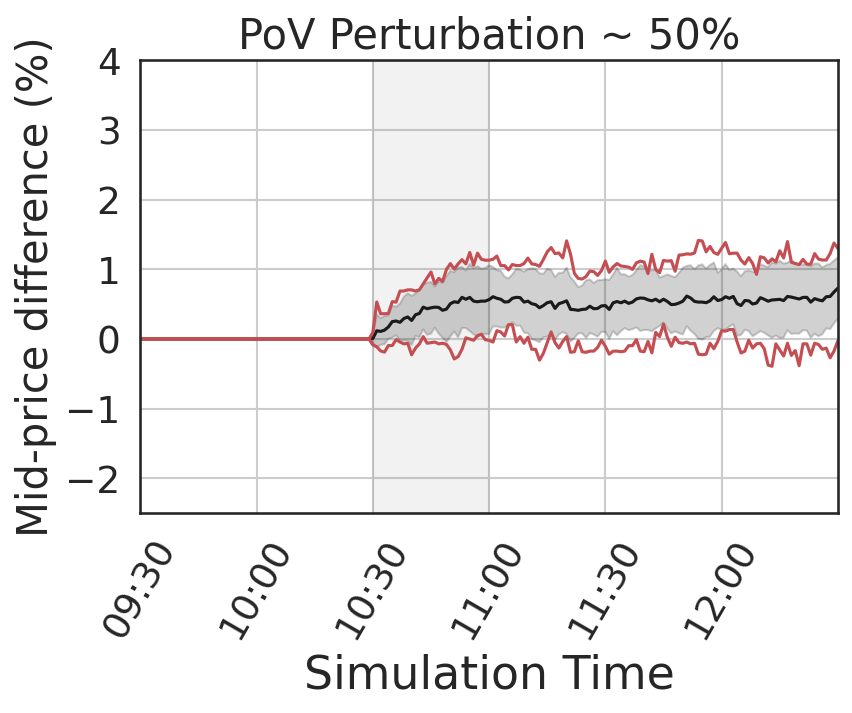} 
  \end{subfigure}  
  \hfill
  \centering
  \begin{subfigure}{0.22\textwidth}
  \includegraphics[width=\textwidth]{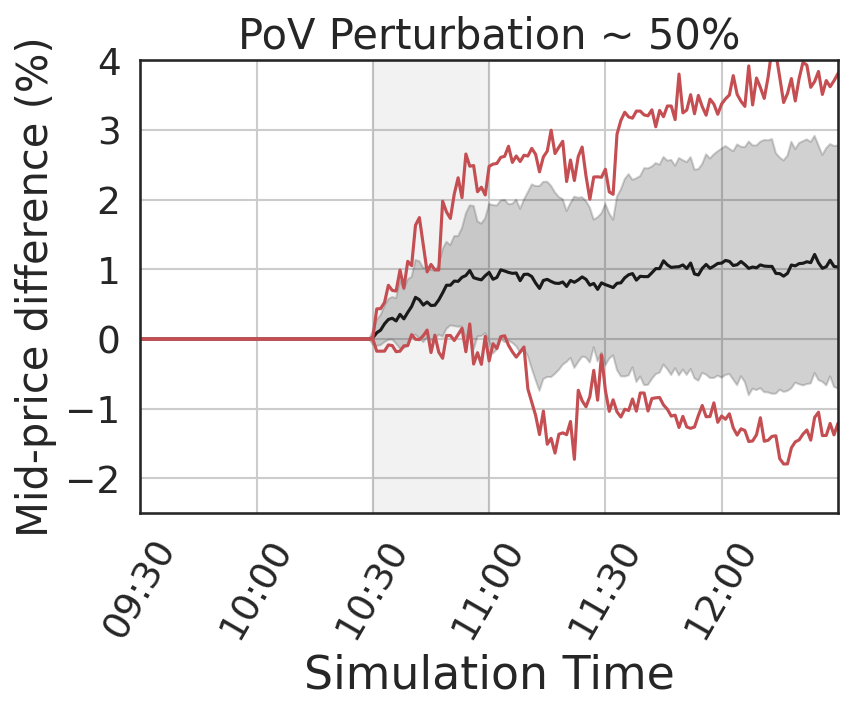}
  \end{subfigure}
  \caption{\small{POV Agent Experiment: explicit model (left) and CGAN-based model (right) exhibit price impact and mean reversion trend.}}\label{fig:pov_impact}
  \vspace{-0.1in}
  \end{figure}

\paragraph{\textbf{Responsiveness}}%
\textit{Responsiveness} to exogenous orders is a desirable property of financial market simulators, and it allows to investigate the impact of a strategy on the market.
To evaluate the responsiveness of our models, we study the price impact caused by an experimental Percent-Of-Volume (POV) agent \cite{balch2019evaluate, coletta2021towards}. This agent submits a burst of buy/sell orders within a limited time window (e.g., 30 minutes) to buy/sell a target amount of shares. This target amount is a percentage $\lambda \in (0, 1]$ of the total transacted volume in the history, for the same time window. 

Figure \ref{fig:pov_impact} shows the simulated market with the $\lambda$-POV agent, with $\lambda \in [0.1, 0.2, 0.5]$. The agent acts only in a time window of $30$ minutes, between 10:30 and 11:00 (gray area in the charts). The charts show the \textit{market impact} as the normalized mid-price difference between the simulation with and without the experimental agent.
The results average 25 different runs: the black line shows the average mid-price difference; the gray shaded region represents one standard deviation; and the red lines represent the 5-th and 95-th percentile. 
The left charts show the explicit model, while the right charts show the CGAN-based model. 

Both models exhibit the price impact, i.e., the burst of buy trades at 10:30 causes prices to rise, showing a substantial deviation w.r.t. the mid-price in the simulation without the experimental agent. 
The greater the value of $\lambda$, the higher the impact on the price. In particular, the average mid-price difference in the CGAN model with $\lambda=0.1$ (top right chart) reaches the 0.5\%, while with $\lambda=0.5$ (bottom right chart) it increases over the 1\%.
With $\lambda \in [0.1, 0.2]$ the CGAN model also shows a mean-reversion to the average price, after the price impact. The mean-reversion effects are weaker for the explicit model, and the price does not return to its average value. With $\lambda = 0.5$ (bottom charts) the experimental agent alters the price trend permanently, i.e., the prices do not return to their average levels. In summary, the observed market impact and price reversion phenomena that arise in simulation, using our world agent approaches, are consistent with observations of the real market \cite{bouchaud2018trades}.

\paragraph{\textbf{Asset returns stylized facts}}
Finally we evaluate the realism of generated time-series against a Multi-Agent Configuration (\textit{MA-Config}), calibrated with 5000 noise, 100 value,
1 market maker, and 25 momentum agents, according to the configurations used in \cite{vyetrenko2019get}. 
The first two charts show the \textit{Minutely Log Returns} and the \textit{Autocorrelation}, respectively, which demonstrate that our models are closest to historical data compared to the multi-agent configuration (i.e., real and synthetic distributions overlap).
The third chart shows the average autocorrelation of square returns as a function of time lag. It decays for both historical and our synthetic data, as time lag increases, while the multi-agent simulator shows an increasing trend. 
We conclude that our models provide a more realistic simulation, compared to the hand-crafted multi-agent configuration.

  \begin{figure}[t]
\centering
  \begin{subfigure}{0.325\linewidth} 
  \includegraphics[width=1\textwidth]{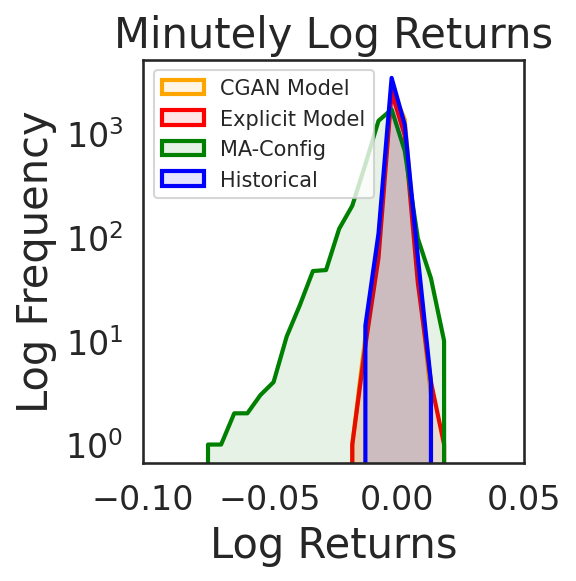} 
  \end{subfigure}
  \hfill
\centering
  \begin{subfigure}{0.325\linewidth}
  \includegraphics[width=1\textwidth]{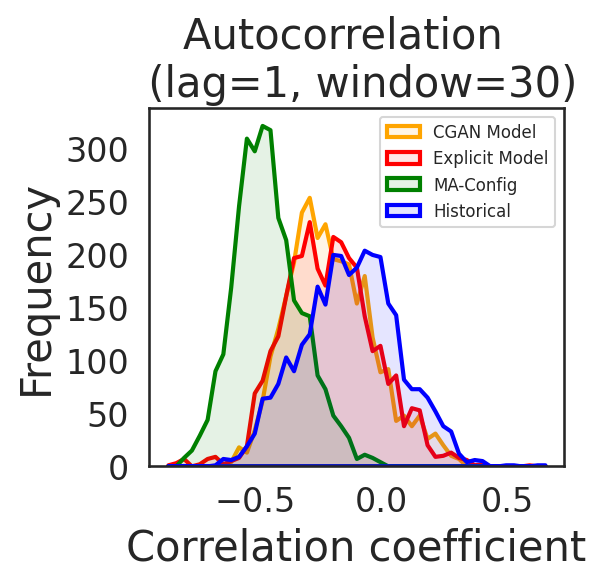}
  \end{subfigure} 
  \hfill
  \centering
  \begin{subfigure}{0.325\linewidth}
  \includegraphics[width=1\textwidth]{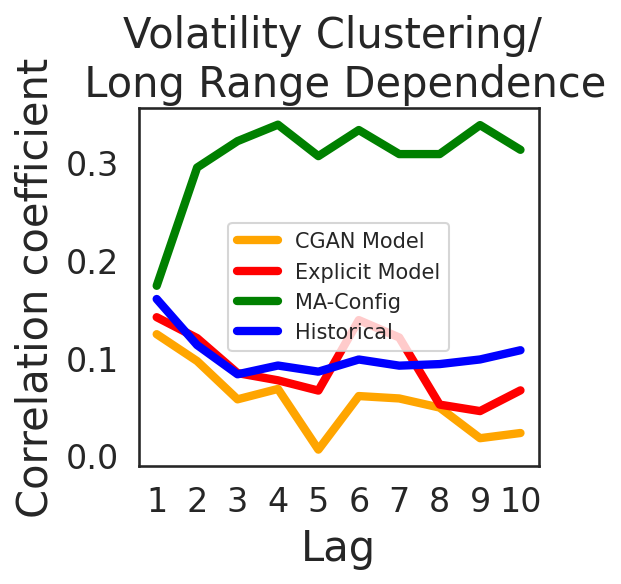}
  \end{subfigure}
  \vspace{-0.1in}
  \caption{\small{Asset returns stylized fact (AVXL): the world models described in the paper produce stylized facts closer to the historical ones when compared to the multi-agent simulation.}}\label{fig:styl_facts}
  \end{figure}

\section{Conclusions}
In this paper we introduced a world model simulator to ease financial simulation, and improve realism and responsiveness.  
We proposed two approaches to the world model based on a Conditional Generative Adversarial Network (CGAN), and a mixture of parametric distributions.
We proved that our world models can learn to simulate realistic markets once trained on historical data, without the need of access to individual and proprietary strategies. 
We also demonstrated that our models improve previous state-of-art solutions, providing more realistic simulation. We discussed the main advantages of the CGAN model, and we demonstrated that it is able to represent a wider range of small-tick stocks. For the future work, we would like to explore and improve the CGAN performance on large-tick stocks, which require a more sophisticated training procedure due to their quasi-degenerate data distributions.
%

\section*{Disclaimer} 
This paper was prepared for informational purposes in part by the Artificial Intelligence Research group of JPMorgan Chase \& Co. and its affiliates (``J.P. Morgan''), and is not a product of the Research Department of J.P. Morgan. J.P. Morgan makes no representation and warranty whatsoever and disclaims all liability, for the completeness, accuracy or reliability of the information contained herein. This document is not intended as investment research or investment advice, or a recommendation, offer or solicitation for the purchase or sale of any security, financial instrument, financial product or service, or to be used in any way for evaluating the merits of participating in any transaction, and shall not constitute a solicitation under any jurisdiction or to any person, if such solicitation under such jurisdiction or to such person would be unlawful.

\bibliographystyle{ACM-Reference-Format}
\bibliography{icaifbib}


\begin{thebibliography}{26}


\ifx \showCODEN    \undefined \def \showCODEN     #1{\unskip}     \fi
\ifx \showDOI      \undefined \def \showDOI       #1{#1}\fi
\ifx \showISBNx    \undefined \def \showISBNx     #1{\unskip}     \fi
\ifx \showISBNxiii \undefined \def \showISBNxiii  #1{\unskip}     \fi
\ifx \showISSN     \undefined \def \showISSN      #1{\unskip}     \fi
\ifx \showLCCN     \undefined \def \showLCCN      #1{\unskip}     \fi
\ifx \shownote     \undefined \def \shownote      #1{#1}          \fi
\ifx \showarticletitle \undefined \def \showarticletitle #1{#1}   \fi
\ifx \showURL      \undefined \def \showURL       {\relax}        \fi
\providecommand\bibfield[2]{#2}
\providecommand\bibinfo[2]{#2}
\providecommand\natexlab[1]{#1}
\providecommand\showeprint[2][]{arXiv:#2}

\bibitem[Balch et~al\mbox{.}(2019)]%
        {balch2019evaluate}
\bibfield{author}{\bibinfo{person}{Tucker~Hybinette Balch},
  \bibinfo{person}{Mahmoud Mahfouz}, \bibinfo{person}{Joshua Lockhart},
  \bibinfo{person}{Maria Hybinette}, {and} \bibinfo{person}{David Byrd}.}
  \bibinfo{year}{2019}\natexlab{}.
\newblock \showarticletitle{How to Evaluate Trading Strategies: Single Agent
  Market Replay or Multiple Agent Interactive Simulation?}
\newblock \bibinfo{journal}{\emph{arXiv preprint arXiv:1906.12010}}
  (\bibinfo{year}{2019}).
\newblock


\bibitem[Bouchaud et~al\mbox{.}(2018)]%
        {bouchaud2018trades}
\bibfield{author}{\bibinfo{person}{Jean-Philippe Bouchaud},
  \bibinfo{person}{Julius Bonart}, \bibinfo{person}{Jonathan Donier}, {and}
  \bibinfo{person}{Martin Gould}.} \bibinfo{year}{2018}\natexlab{}.
\newblock \bibinfo{booktitle}{\emph{Trades, quotes and prices: financial
  markets under the microscope}}.
\newblock \bibinfo{publisher}{Cambridge University Press}.
\newblock


\bibitem[Byrd et~al\mbox{.}(2020)]%
        {byrd2020abides}
\bibfield{author}{\bibinfo{person}{David Byrd}, \bibinfo{person}{Maria
  Hybinette}, {and} \bibinfo{person}{Tucker~Hybinette Balch}.}
  \bibinfo{year}{2020}\natexlab{}.
\newblock \showarticletitle{ABIDES: Towards High-Fidelity Multi-Agent Market
  Simulation}. In \bibinfo{booktitle}{\emph{Proceedings of the 2020 ACM SIGSIM
  Conference on Principles of Advanced Discrete Simulation}}.
  \bibinfo{pages}{11--22}.
\newblock


\bibitem[Chiarella and Iori(2002)]%
        {chiarella2002simulation}
\bibfield{author}{\bibinfo{person}{Carl Chiarella} {and}
  \bibinfo{person}{Giulia Iori}.} \bibinfo{year}{2002}\natexlab{}.
\newblock \showarticletitle{A simulation analysis of the microstructure of
  double auction markets}.
\newblock \bibinfo{journal}{\emph{Quantitative finance}} \bibinfo{volume}{2},
  \bibinfo{number}{5} (\bibinfo{year}{2002}), \bibinfo{pages}{346}.
\newblock


\bibitem[Cho and Norman(2021)]%
        {cho2021bit}
\bibfield{author}{\bibinfo{person}{Christopher~J Cho} {and}
  \bibinfo{person}{Timothy~J Norman}.} \bibinfo{year}{2021}\natexlab{}.
\newblock \showarticletitle{Bit by bit: how to realistically simulate a
  crypto-exchange}. In \bibinfo{booktitle}{\emph{Proceedings of the Second ACM
  International Conference on AI in Finance}}. \bibinfo{pages}{1--9}.
\newblock


\bibitem[Coletta et~al\mbox{.}(2021)]%
        {coletta2021towards}
\bibfield{author}{\bibinfo{person}{Andrea Coletta}, \bibinfo{person}{Matteo
  Prata}, \bibinfo{person}{Michele Conti}, \bibinfo{person}{Emanuele Mercanti},
  \bibinfo{person}{Novella Bartolini}, \bibinfo{person}{Aymeric Moulin},
  \bibinfo{person}{Svitlana Vyetrenko}, {and} \bibinfo{person}{Tucker Balch}.}
  \bibinfo{year}{2021}\natexlab{}.
\newblock \showarticletitle{Towards Realistic Market Simulations: A Generative
  Adversarial Networks Approach}. In \bibinfo{booktitle}{\emph{Proceedings of
  the Second ACM International Conference on AI in Finance (ICAIF)}}.
\newblock


\bibitem[Cont(2001)]%
        {cont2001empirical}
\bibfield{author}{\bibinfo{person}{Rama Cont}.}
  \bibinfo{year}{2001}\natexlab{}.
\newblock \showarticletitle{Empirical properties of asset returns: stylized
  facts and statistical issues}.
\newblock \bibinfo{journal}{\emph{Quantitative finance}} \bibinfo{volume}{1},
  \bibinfo{number}{2} (\bibinfo{year}{2001}), \bibinfo{pages}{223}.
\newblock


\bibitem[Farmer and Foley(2009)]%
        {farmer2009economy}
\bibfield{author}{\bibinfo{person}{J~Doyne Farmer} {and}
  \bibinfo{person}{Duncan Foley}.} \bibinfo{year}{2009}\natexlab{}.
\newblock \showarticletitle{The economy needs agent-based modelling}.
\newblock \bibinfo{journal}{\emph{Nature}} \bibinfo{volume}{460},
  \bibinfo{number}{7256} (\bibinfo{year}{2009}), \bibinfo{pages}{685--686}.
\newblock


\bibitem[Farmer et~al\mbox{.}(2005)]%
        {farmer2005predictive}
\bibfield{author}{\bibinfo{person}{J~Doyne Farmer}, \bibinfo{person}{Paolo
  Patelli}, {and} \bibinfo{person}{Ilija~I Zovko}.}
  \bibinfo{year}{2005}\natexlab{}.
\newblock \showarticletitle{The predictive power of zero intelligence in
  financial markets}.
\newblock \bibinfo{journal}{\emph{Proceedings of the National Academy of
  Sciences}} \bibinfo{volume}{102}, \bibinfo{number}{6} (\bibinfo{year}{2005}),
  \bibinfo{pages}{2254--2259}.
\newblock


\bibitem[Goodfellow(2016)]%
        {goodfellow2016nips}
\bibfield{author}{\bibinfo{person}{Ian Goodfellow}.}
  \bibinfo{year}{2016}\natexlab{}.
\newblock \showarticletitle{Nips 2016 tutorial: Generative adversarial
  networks}.
\newblock \bibinfo{journal}{\emph{arXiv preprint arXiv:1701.00160}}
  (\bibinfo{year}{2016}).
\newblock


\bibitem[Goodfellow et~al\mbox{.}(2014)]%
        {NIPS2014_gan}
\bibfield{author}{\bibinfo{person}{Ian Goodfellow}, \bibinfo{person}{Jean
  Pouget-Abadie}, \bibinfo{person}{Mehdi Mirza}, \bibinfo{person}{Bing Xu},
  \bibinfo{person}{David Warde-Farley}, \bibinfo{person}{Sherjil Ozair},
  \bibinfo{person}{Aaron Courville}, {and} \bibinfo{person}{Yoshua Bengio}.}
  \bibinfo{year}{2014}\natexlab{}.
\newblock \showarticletitle{Generative Adversarial Nets}. In
  \bibinfo{booktitle}{\emph{Advances in Neural Information Processing
  Systems}}, Vol.~\bibinfo{volume}{27}.
\newblock


\bibitem[Gulrajani et~al\mbox{.}(2017)]%
        {gulrajani2017improved}
\bibfield{author}{\bibinfo{person}{Ishaan Gulrajani}, \bibinfo{person}{Faruk
  Ahmed}, \bibinfo{person}{Martin Arjovsky}, \bibinfo{person}{Vincent
  Dumoulin}, {and} \bibinfo{person}{Aaron Courville}.}
  \bibinfo{year}{2017}\natexlab{}.
\newblock \showarticletitle{Improved training of wasserstein gans}.
\newblock \bibinfo{journal}{\emph{arXiv preprint arXiv:1704.00028}}
  (\bibinfo{year}{2017}).
\newblock


\bibitem[Inc.(2020)]%
        {nasdaq2020nasdaq}
\bibfield{author}{\bibinfo{person}{Nasdaq Inc.}}
  \bibinfo{year}{2020}\natexlab{}.
\newblock \bibinfo{title}{NASDAQ TotalView-ITCH 5.0}.
\newblock
\newblock
\urldef\tempurl%
\url{https://www.nasdaqtrader.com/content/technicalsupport/specifications/dataproducts/NQTVITCHSpecification.pdf}
\showURL{%
\tempurl}


\bibitem[LeBaron and Yamamoto(2007)]%
        {lebaron2007long}
\bibfield{author}{\bibinfo{person}{Blake LeBaron} {and}
  \bibinfo{person}{Ryuichi Yamamoto}.} \bibinfo{year}{2007}\natexlab{}.
\newblock \showarticletitle{Long-memory in an order-driven market}.
\newblock \bibinfo{journal}{\emph{Physica A: Statistical mechanics and its
  Applications}} \bibinfo{volume}{383}, \bibinfo{number}{1}
  (\bibinfo{year}{2007}), \bibinfo{pages}{85--89}.
\newblock


\bibitem[Li et~al\mbox{.}(2020)]%
        {li2020generating}
\bibfield{author}{\bibinfo{person}{Junyi Li}, \bibinfo{person}{Xintong Wang},
  \bibinfo{person}{Yaoyang Lin}, \bibinfo{person}{Arunesh Sinha}, {and}
  \bibinfo{person}{Michael Wellman}.} \bibinfo{year}{2020}\natexlab{}.
\newblock \showarticletitle{Generating realistic stock market order streams}.
  In \bibinfo{booktitle}{\emph{Proceedings of the AAAI Conference on Artificial
  Intelligence}}, Vol.~\bibinfo{volume}{34}. \bibinfo{pages}{727--734}.
\newblock


\bibitem[Lo(2005)]%
        {lo2005reconciling}
\bibfield{author}{\bibinfo{person}{Andrew~W Lo}.}
  \bibinfo{year}{2005}\natexlab{}.
\newblock \showarticletitle{Reconciling efficient markets with behavioral
  finance: the adaptive markets hypothesis}.
\newblock \bibinfo{journal}{\emph{Journal of investment consulting}}
  \bibinfo{volume}{7}, \bibinfo{number}{2} (\bibinfo{year}{2005}),
  \bibinfo{pages}{21--44}.
\newblock


\bibitem[Mirza and Osindero(2014)]%
        {mirza2014conditional}
\bibfield{author}{\bibinfo{person}{Mehdi Mirza} {and} \bibinfo{person}{Simon
  Osindero}.} \bibinfo{year}{2014}\natexlab{}.
\newblock \bibinfo{title}{Conditional Generative Adversarial Nets}.
\newblock
\newblock
\showeprint[arxiv]{1411.1784}~[cs.LG]


\bibitem[Mizuta(2016)]%
        {mizuta2016brief}
\bibfield{author}{\bibinfo{person}{Takanobu Mizuta}.}
  \bibinfo{year}{2016}\natexlab{}.
\newblock \showarticletitle{A brief review of recent artificial market
  simulation (agent-based model) studies for financial market regulations
  and/or rules}.
\newblock \bibinfo{journal}{\emph{Available at SSRN 2710495}}
  (\bibinfo{year}{2016}).
\newblock


\bibitem[Mizuta(2020)]%
        {mizuta2020agent}
\bibfield{author}{\bibinfo{person}{Takanobu Mizuta}.}
  \bibinfo{year}{2020}\natexlab{}.
\newblock \showarticletitle{An agent-based model for designing a financial
  market that works well}. In \bibinfo{booktitle}{\emph{2020 IEEE symposium
  series on computational intelligence (SSCI)}}. IEEE,
  \bibinfo{pages}{400--406}.
\newblock


\bibitem[NASDAQ({[n.\,d.]})]%
        {totalview}
\bibfield{author}{\bibinfo{person}{NASDAQ}.}
  \bibinfo{year}{[n.\,d.]}\natexlab{}.
\newblock \bibinfo{title}{Nasdaq Total View}.
\newblock
\newblock
\urldef\tempurl%
\url{https://www.nasdaq.com/solutions/nasdaq-totalview}
\showURL{%
\tempurl}


\bibitem[Palit et~al\mbox{.}(2012)]%
        {palit2012can}
\bibfield{author}{\bibinfo{person}{Imon Palit}, \bibinfo{person}{Steve Phelps},
  {and} \bibinfo{person}{Wing~Lon Ng}.} \bibinfo{year}{2012}\natexlab{}.
\newblock \showarticletitle{Can a zero-intelligence plus model explain the
  stylized facts of financial time series data?}. In
  \bibinfo{booktitle}{\emph{Proceedings of the 11th International Conference on
  Autonomous Agents and Multiagent Systems-Volume 2}}.
  \bibinfo{pages}{653--660}.
\newblock


\bibitem[Paulin et~al\mbox{.}(2018)]%
        {paulin2018agent}
\bibfield{author}{\bibinfo{person}{James Paulin}, \bibinfo{person}{Anisoara
  Calinescu}, {and} \bibinfo{person}{Michael Wooldridge}.}
  \bibinfo{year}{2018}\natexlab{}.
\newblock \showarticletitle{Agent-based modeling for complex financial
  systems}.
\newblock \bibinfo{journal}{\emph{IEEE Intelligent Systems}}
  \bibinfo{volume}{33}, \bibinfo{number}{2} (\bibinfo{year}{2018}),
  \bibinfo{pages}{74--82}.
\newblock


\bibitem[Raberto et~al\mbox{.}(2001)]%
        {raberto2001agent}
\bibfield{author}{\bibinfo{person}{Marco Raberto}, \bibinfo{person}{Silvano
  Cincotti}, \bibinfo{person}{Sergio~M Focardi}, {and} \bibinfo{person}{Michele
  Marchesi}.} \bibinfo{year}{2001}\natexlab{}.
\newblock \showarticletitle{Agent-based simulation of a financial market}.
\newblock \bibinfo{journal}{\emph{Physica A: Statistical Mechanics and its
  Applications}} \bibinfo{volume}{299}, \bibinfo{number}{1-2}
  (\bibinfo{year}{2001}), \bibinfo{pages}{319--327}.
\newblock


\bibitem[Vyetrenko et~al\mbox{.}(2020)]%
        {vyetrenko2019get}
\bibfield{author}{\bibinfo{person}{Svitlana Vyetrenko}, \bibinfo{person}{David
  Byrd}, \bibinfo{person}{Nick Petosa}, \bibinfo{person}{Mahmoud Mahfouz},
  \bibinfo{person}{Danial Dervovic}, \bibinfo{person}{Manuela Veloso}, {and}
  \bibinfo{person}{Tucker~Hybinette Balch}.} \bibinfo{year}{2020}\natexlab{}.
\newblock \showarticletitle{Get Real: Realism Metrics for Robust Limit Order
  Book Market Simulations}. In \bibinfo{booktitle}{\emph{ACM International
  Conference on AI in Finance (ICAIF)}}.
\newblock


\bibitem[Wang et~al\mbox{.}(2017)]%
        {wang2017stockyard}
\bibfield{author}{\bibinfo{person}{Jianling Wang}, \bibinfo{person}{Vivek
  George}, \bibinfo{person}{Tucker Balch}, {and} \bibinfo{person}{Maria
  Hybinette}.} \bibinfo{year}{2017}\natexlab{}.
\newblock \showarticletitle{Stockyard: A discrete event-based stock market
  exchange simulator}. In \bibinfo{booktitle}{\emph{2017 Winter Simulation
  Conference (WSC)}}. IEEE, \bibinfo{pages}{1193--1203}.
\newblock


\bibitem[Wang and Wellman(2017)]%
        {wang2017spoofing}
\bibfield{author}{\bibinfo{person}{Xintong Wang} {and}
  \bibinfo{person}{Michael~P Wellman}.} \bibinfo{year}{2017}\natexlab{}.
\newblock \showarticletitle{Spoofing the Limit Order Book: An Agent-Based
  Model}. In \bibinfo{booktitle}{\emph{Proceedings of the 16th Conference on
  Autonomous Agents and MultiAgent Systems}}. \bibinfo{pages}{651--659}.
\newblock


\end{thebibliography}

\end{document}